\documentclass[a4paper,fleqn,usenatbib]{mnras}

\usepackage{newtxtext,newtxmath}

\usepackage[T1]{fontenc}
\usepackage{ae,aecompl}

\usepackage{graphics}	
\usepackage{epstopdf}
\usepackage{amsmath}	
\usepackage{amssymb}	
\usepackage{pdflscape}
\usepackage{rotating}
\usepackage{threeparttable}
\usepackage{adjustbox}

\title[Specific angular momenta of superthin galaxies]
{The specific angular momenta of superthin galaxies: Cue to their origin?}

\author[V. Jadhav Y \& A. Banerjee]
       {Vikas Jadhav Y$^{1}$,
        and Arunima Banerjee$^{2}$\thanks{E-mail : arunima@iisertirupati.ac.in}  \\
 $^1$, $^2$  Indian Institute of Science Education and Research, Tirupati 517507, India \\
}

\begin{document}
\label{firstpage}
\pagerange{\pageref{firstpage}--\pageref{lastpage}}
\maketitle
\begin{abstract}

Superthin galaxies are low surface brightness (LSB) bulgeless disc galaxies having stellar discs with unusually high planar-to-vertical axes ratio $b/a$ $>$ 10 - 20, the formation and evolution of which is not well-understood. We calculate the specific angular momenta of a sample of six superthins and nine other bulgeless LSBs using stellar photometry, atomic hydrogen (HI) surface density and high resolution HI rotation curves available in the literature. We find that the stellar specific angular mometum $j_{\rm{s}}$, and hence the stellar disc size given by the exponential stellar disc scale length R$_D$, of three superthins and seven LSBs lie above the 95.4 $\%$ confidence band of the $j_{\rm{s}}$ - $V_{\rm{rot}}$ regression line for ordinary bulgeless disc galaxies, $V_{\rm{rot}}$ being the asymptotic rotational velocity. Further, we find that superthins and LSBs have higher $j_{\rm{s}}$ and R$_D$ values for a given value of stellar mass $M_{\rm{s}}$ at high values of statistical significance, compared to ordinary disc galaxies. Therefore, we conclude, a superthin is may be distinguished by a characteristically larger disc size which could possibly explain the origin of its large planar-to-vertical axes ratio. Interestingly, we find that the median spin parameter $\lambda = \frac{ j_{\rm{stars}} } { {\sqrt(2) V_{\rm{vir}} R_{\rm{vir}} } }$, $V_{\rm{vir}}$ and $R_{\rm{vir}}$ being the virial velocity and virial radius of the galaxy respectively,  is 0.13 $\pm$ 0.01 for superthin galaxies which is an order of magnitude higher than those of LSBs and ordinary disc galaxies, which may have important implications for the existence of superthin stellar discs in these low surface brightness galaxies.

\end{abstract}


\section{Introduction}

Superthin galaxies are low surface brightness (LSB), flat or bulgeless disc galaxies with strikingly high values of planar-to-vertical axes ratio $b/a \sim 10 - 20$,
also referred to as ultra-flat or very thin galaxies (Goad \& Roberts 1981, Karachentsev 1989; Matthews, van Driel \& Gallagher 1999, Karachentsev et al. 2016, Bizyaev et al. 2017). They are gas-rich and dark matter-dominated with low metallicities and star formation rates (See Kautsch 2009 for a review). Superthins therefore serve as proxies of a very late-type galaxy population and are therefore ideal laboratories for testing models of galaxy evolution in the local universe.

A superthin stellar vertical structure implies an ultra-cold stellar disc i.e., a stellar disc with distinctly low values of stellar dispersion in the vertical 
direction. The existence of ultra-cold stellar discs is a mystery, given the hierarchical structure formation scenario of the current cosmological paradigm. 
Besides, galaxy mergers are known to be common in group environments, which results in the heating and subsequent thickening of the stellar discs (Walker, Mihos \& Hernquist 1996;
Velazquez \& White 1999; Qu et al. 2011). LSBs are however favourably located at the outskirts of voids, which, therefore, complies with the quiescent nature of their stellar discs (Rosenbaum et al. 2009).
However, galactic discs may also be heated by internal mechanisms via the growth of non-axisymmetric instabilities like bars, spiral arms and other mediators of 
secular evolution (Barbanis \& Woltjer 1967, Grand et al. 2016). This possibly implies that superthins are devoid of strong bars and spiral arms, which is characteristic of any very late-type, under-evolved
systems. Observational confirmation of the absence or near-absence of bars and spiral arms in highly edge-on systems like the superthins is tricky (But see, Bureau
 \& Athanoussoula 2005). However, Zasov, Makarov \& Mikailhova (1991) showed that superthin galaxies are stabilised against vertical bending instabilities by a massive dark matter halos (See, also, Khoperskov et al. (2010)). 
 Analytical calculations by Ghosh \& Jog (2014) also indicated that the growth of both axi-symmetric and non-axisymmetric instabilities in the superthin galaxy UGC7321 are suppressed by its dark matter halo. 
Garg \& Banerjee (2017) and Narayanan \& Banerjee (2019, in preparation) demonstrated that the dark matter suppresses the growth of both axi-symmetric and 
non axi-symmetric instabilities for a sample of LSBs in general (Also, see, Banerjee \& Jog 2013, Narayanan \& Banerjee 2018).  Therefore, the heating of the stellar disc via secular evolution
 is possibly inhibited by the dark matter halo in superthin galaxies, which is a viable reason for the survival of their razor-thin stellar discs.
 
 Alternatively, a higher value of the stellar specific angular momentum in a given radial gravitational field should favour the formation of a larger stellar disc, which may possibly get reflected in a larger planar-to-vertical axes ratio of the same. In fact, from an empirical study of the size-mass data from the SDSS, Shen et al. (2013) found that late-type galaxies, on an average, have larger characteristic size compared to their early-type counterparts.
The structure and morphology of a disc galaxy is primarily regulated by a counterbalance between the inward radial gravitational field and the outward centrifugal force, in contrast to ellipticals or spheroidals in which the dynamics is predominantly driven by the velocity dispersion of the stars. In fact, as a physical rationale behind the galaxy classification scheme, Fall (1983) proposed the  
 \emph{j$_s$ - M$_s$} where j$_s$ is the specific angular momentum of the stars and M$_s$ the stellar mass, often referred to as the Fall relation, according to which, galaxies with M$_s$ $\ge$ 9 obey the
  j$_s$ - M${_s}^{2/3}$ scaling relation (Also see Romanowsky \& Fall 2012). Interestingly, the specific angular momentum of the dark matter halo also scales as their mass to the power 2/3 according to the 
  tidal torque theory, which assumes that the dark matter halos acquire their angular momenta via gravitational torques during the early phases of formation and evolution (Peebles 1969; White 1984; Barnes \& Efstathiou 
 1987). According to the modern theory of galaxy formation and evolution, galaxies form via the cooling and condensation of baryons in the potential wells of their host dark matter halos (Rees \& White 1979) and therefore the 
 Fall relation indicates the intrinsic connection between the galactic disc and its host dark matter halo. In fact, numerical simulations have indicated that gas and dark matter within virialized systems have comparable values of angular momenta (See, for example, Sharma \& Steinmetz 2005), and the assumption that the stellar disc originating from star formation in the cold gas disc acquires the same angular momentum as the parent gas disc is in agreement with the Fall relation. However, the connection is primarily interceded by two fundamental parameters, namely the global star-formation efficiency defined as the ratio of the stellar mass to halo mass, and the retained fraction of the specific angular momentum which is the ratio of the specific angular momentum of the stars to that of the dark matter halo, which, in turn, are governed by complex process of galaxy formation and evolution including stellar feedback,
 gas dissipation and exchange of angular momentum among the different baryonic components.

 In this paper, we study the specific angular momenta of a sample of six superthin galaxies to understand the origin of their characteristic razor-thin stellar discs i.e., stellar discs with unusually high planar-to-vertical axes ratios,
 for which stellar photometry, HI surface density and high resolution HI rotation curves were already available in the literature. However, the angular momenta of their discs were not studied earlier.
 In addition, we also consider a sample of nine general low surface brightness galaxies (LSBs)  for which again all the necessary input parameters were already available in the literature as well.
 None of our sample LSBs are edge-on and hence we have no notion if their stellar discs are superthin or not. However, they were part of the sample of galaxies considered for obtaining the fundamental 
 angular momentum-mass relation encompassing all galaxy morphological types from dwarf-irregulars to massive spirals (Posti et al. 2018). We primarily check if our sample superthins and LSBs have statistically higher values of specific angular momenta of their stellar discs j$_{\rm{s}}$ for a given value of (i) the radial gravitational field as indicated by their asymptotic rotational velocity $v_{\rm{rot}}$ and (ii) the stellar mass  M$_s$, as compared to ordinary bulgeless disc galaxies chosen from the sample of Obreschkow \& Glazebrook (2014).  Since the gas discs are progenitors of the stellar discs, we repeat the same study for the gas as well as the total (stars + gas) baryonic disc to gain possible insights into the formation and evolution processes of superthin galaxies.  In addition,
 we study the dependence of $R_D$ on the properties of its host dark matter halo as was of late investigated in hydrodynamical cosmological simulation studies. Finally, we also study the dependence of  j$_{\rm{star}}$ on the gas mass fraction M$_{\rm{g}}$/{M$_{\rm{b}}$} where M$_{\rm{g}}$ and {M$_{\rm{b}}$} are the gas mass and total baryonic mass respectively as predicted by recent semi-analytical studies of galaxy formation and evolution.

 The rest of the paper is organized as follows: In \S 2, we discuss the theory and numerical calculations, in \S 3 the sample, in \S 4 the input parameters followed by results and discussion in \S 5 and conclusions in \S 6. 

\section{Data Analysis}
\subsection{Theory \& Numerical Calculations}

The angular momentum of the $i^{\rm{th}}$ component of the disc for axis-symmetric discs is given by
$$ J_i = 2 \pi \int_{0}^{\infty} {\Sigma}_i (\rm{R}) v(\rm{R}) \rm{R} \rm{dR}  \eqno(1) $$
where $\Sigma (\rm{R})$ is the radial surface density profile of the $i^{\rm{th}}$ component and $v(\rm{R})$ rotational velocity. The mass of 
the $i^{\rm{th}}$ component is given by
$$ M_i = 2 \pi \int_{0}^{\infty} {\Sigma}_i (\rm{R}) \rm{R} \rm{dR} \eqno(2) $$ 
Therefore, the specific angular momentum is given by 
$$ j_i = \frac{J_i}{M_i} \eqno(3) $$

Earlier studies have shown that in the optical i.e., B-band or R-band, the surface brightness profile is well-fitted with an exponential (See, for example, de Blok et al. 2001), whereas in the mid infra-red, the same is fitted well with a double exponential (See, for example, Salo et al. 2015). The same trend holds good for our sample of superthins and LSBs. Therefore, for our sample stellar discs, ${\Sigma}_i (\rm{R})$ is either an exponential 
$${\Sigma}_s (R) = {\Sigma}_s (0) \rm {exp} (-R/R_D) \eqno(4a)$$ 
where ${\Sigma}_s (0)$ is the central stellar surface density and $R_D$ the exponential stellar disc scalelength.
or, a double exponential given by
$${\Sigma}_s (R) = {\Sigma}_s (0,1) \rm{exp} (-R/R_D(1)) + {\Sigma}_s (0,2) exp(-R/R_D(2)) \eqno(4b)$$
where ${\Sigma}_s (0,1)$ is the central stellar surface density and $R_D(1)$ the exponential stellar disc scalelength of stellar disc 1 and so on.\\ \\

Similarly, earlier work indicated that the radial profiles of HI surface density could be well-fitted with double-gaussians profiles (See, for example, Begum \& Chengalur 2004, Patra et al. 2014), possibly signifying the presence of two HI discs. Also, galaxies with the HI surface density peaking away from the centre are common, which indicates the presence of an HI hole at the centre. Our sample HI surface density profiles could therefore be fitted well with off-centred double Gaussians given by

$$ {\Sigma}_g (R) = {\Sigma}_g (0,1) \rm{exp} [-{\frac{{(r-a_1)}^2}{2 {r_{0,1}}^2}}] + {\Sigma}_g (0,2) \rm{exp} [-{\frac{{(r-a_2)}^2}{2 {r_{0,2}}^2}}] \eqno(5)$$
where ${\Sigma}_g (0, 1)$ is the central gas surface density, $a_1$ the centre and $r_{0,1}$ the scalelength of gas disc 1 and so on. For the gas disc, we consiser the atomic hydrogen (HI) surface density only as the presence of molecular gas in LSBs is known to be negligible (See, for example, Banerjee \& Bapat for a discussion).\\ \\ 
Finally, the rotation curves of galaxies in general are commonly well-fitted either by a Brand-profile (See, for example, Banerjee \& Bapat 2014)  given by 

$$ v(R) = \frac{ V_{\rm{rot}} R/R_{\rm{max}}}{ {(1/3 + 2/3 {(R/R_{\rm{max}})}^n)}^{3/2}}  \eqno(6a)$$
where $V_{\rm{max}} $, $R_{\rm{max}}$ and $n$ are the free parameters, with $V_{\rm{rot}} $ the asymptotic rotational velocity. \\
or, with an exponential profile given by 
$$ v(R) = \alpha (1 - {\rm{exp}}(-R/\beta) ) \eqno(6b)$$
where $\alpha$ and $\beta$ are free parameters (See, for example, Obreschkow \& Glazebrook 2014). \\

\noindent \textbf{Error bars:} The error bars on the total angular momentum ($J_s$, $J_g$, $J_b$), total mass ($M_s$, $M_g$, $M_b$) and the total specific angular momentum ($j_s$, $j_g$, $j_b$) were calculated by propagating the fitting errors on the input parameters.

\subsection{Sample} Our sample consists of six superthin galaxies for which high resolution rotation curves and gas surface density profiles from 
HI 21cm radio-synthesis observations were available in the literature: UGC7321, IC5249, IC2233, UGC711, NGC4244 and FGC1540.  We may note here that two of our sample superthin galaxies, UGC7321 and IC2233, are from the original sample of classic superthin galaxies, studied by Goad \& Roberts (1981) although IC2233 has a major-to-minor axes ratio $b/a$ $<$ 10. Our sample superthins are moderately inclined to perfectly edge-on with angles of inclination $i$ ranging between 65$^o$ - 90$^o$, with unusually high values of  $b/a$ 
$\sim$ 9 - 16, $B$ band central surface brightness $\sim$ 22.4 -24.5 mag arcsec$^{-2}$ and low-to-intermediate values of the asymptotic rotational velocities v$_{\rm{max}}$ $\sim$ 85-112 kms$^{-1}$.   \\

\noindent The basic properties of our sample superthins in the $B$-band and 3.6 $\mu$ band are summarized in Table 1 and 2 respectively. $B$-band
 photometry for FGC1540 was not available in the literature and hence has not been presented in the table. The $B$-band, the stellar disc appears to be superthin, however, traces the young stellar population only. The major mass fraction of the stellar disc is however trace
  by the mid infra-red band of 3.6$mu$. Unlike in the $B$-band, in the 3.6$mu$ band, for four out of our six superthins, the stars are distributed in two exponential discs with different radial scalelengths and vertical scaleheights, with one of the discs (Disc 1) several times more massive than the other (Disc 2) in general. We also note that Disc 1 is fainter i.e., with a lower surface brightness and with a larger disc size. \\

\begin{table}
	\centering
	\begin{minipage}{150mm}
	\caption{Basic properties of our sample of superthin galaxies}
	\label{tab:example_table}
	\begin{tabular}{lcccccc} 
		\hline
		Name\footnote{All the quantities except for Distance and Central B-band \\ surface brightness have been quoted from NED/Hyperleda}& D\footnote{Distance from Tully et al. (2013)} & ${{\mu}_B}(0)$\footnote{B-band central surface brightness} & $b/a$\footnote{Major-to-minor axes ratio} & $i$\footnote{Inclination} & $V_{\rm{rot}}$\footnote{Maximum Velocity} & $R_{\rm{d}}$\footnote{Radial disc scale length}\\ 
                     &(Mpc) & (mag arcsec$^{-2}$) & (kpc)& $^o$ & (kms$^{-1}$) & (kPc) \\
		\hline
		UGC7321 & 22.18 & 23.5 & 15.7  & 90 & 103 & 4.6\\
		IC5249 & 31.77 & 24.5  & 10.2 & 90 &108 & 7.0\\
		IC2233 & 12.59& 22.6 & 8.9& 90 &110 & 1.5\\
		UGC711 & 20.61& 23.6  &15.5 &74.7 &105& 3.2  \\
		NGC4244 &4.62 & 22.4   &9.1& 65.4 & 97 & 2.0\\
		FGC1540 & 20.32& &  12.25& 78.7& 83&  \\		
		\hline
	\end{tabular}
\end{minipage}
\end{table}

\begin{table}
	\centering
	\begin{minipage}{150mm}
	\caption{GALFIT structural decompositions of the superthin galaxies in the 3.6 $\micron$ band}
	\label{tab:example_table}
	\begin{tabular}{lcccc} 
		\hline
		Name\footnote{Adapted from Salo et al. (2015)} & ${{\mu}_0}(1)$\footnote{Central surface brightness of outer disc} & ${{R}_{d}}(1)$\footnote{Radial scale length of outer disk} & ${{\mu}_0}(2)$\footnote{Central surface brightness of inner disc} & ${{R}_{d}}(2)$\footnote{Radial scale length of inner disk} \\ 
		& (mag arcsec$^{-2}$)& (kpc) & (mag arcsec$^{-2}$) & (kpc)\\
		\hline
		UGC7321 & 21.73 & 5.26 & 19.94 & 2.2 \\
		IC5249  & 21.70 & 5.24 & 20.53 & 1.23 \\
		IC2233  & 21.67 & 2.16 & 20.82 & 0.81\\
		UGC711  &  21.189     & 2.14 &  -   &  -   \\
		NGC4244 & 20.27 & 1.62 &    -   &  -    \\
		FGC1540 & 22.23 & 1.87 & 21.39 & 0.54 \\

		\hline
	\end{tabular}
\end{minipage}
\end{table}	

\noindent In addition, we choose a sample of nine bulgeless LSBs for which $B$-band stellar photometry, HI surface density and high resolution HI rotation curves were already available 
(de Blok et al. 2001).  Our sample LSBs are nearly to moderately face-on with angles of inclination $i$ ranging between 25$^o$ - 65$^o$, and therefore we cannot assess if they are superthin. Their $B$ band central surface brightness ${\Sigma}_{s}$(0) $\sim$ 22.10 -24.03 mag arcsec$^{-2}$, asymptotic rotational velocities v$_{\rm{max}}$ $\sim$ 70-142 kms$^{-1}$ and therefore have almost comparable photometric and dynamical properties as the superthins. The basic properties of our sample LSBs in the B-band are given in Table 3. The rotation curves of our sample superthins are found to be well-fitted by a Brandt profile whereas the LSBs with an exponential profile (See \S 2). 
GALFIT 2-D structural decompositions in the 3.6$mu$ band were not available for our sample of LSBs. \\

\begin{table}
        \centering
        \begin{minipage}{150mm}
        \caption{Basic properties of our sample of LSBs}
        \label{tab:example_table}
        \begin{tabular}{lccccc} 
                \hline
                Name\footnote{All values quoted from de Blok et al. (2001)} & D\footnote{Distance} & ${{\mu}_B}(0)$\footnote{B-band central surface brightness} & $R_D$\footnote{Exponential disc scalelength} & i\footnote{Inclination} & $V_{\rm{max}}$\footnote{Maximum Velocity}\\
                     &(Mpc) & (mag arcsec$^{-2}$) & (kpc)& ($^o$) & (kms$^{-1}$) \\
                \hline
				     
                $F563-V2 $&$46 $&$22.10 $&$ 2.1 $&$29^a  $&$118$ \\
                $F574-1  $&$72 $&$23.31 $&$ 4.3 $&$65    $&$100$ \\            
                $F583-1  $&$24 $&$24.03 $&$ 1.6 $&$63    $&$87 $ \\
                $F583-4  $&$37 $&$23.76 $&$ 2.7 $&$55    $&$70 $ \\
		$F563-1  $&$45 $&$23.6  $&$ 2.8 $&$25    $&$112$ \\
		$F568-V1 $&$80 $&$23.3  $&$ 3.2 $&$40    $&$118$ \\
		$F568-1  $&$85 $&$23.8  $&$ 5.3 $&$26    $&$142$ \\
		$F568-3  $&$77 $&$23.1  $&$ 4.0 $&$40    $&$105$ \\
		$F579-V1 $&$85 $&$22.8  $&$ 5.1 $&$26    $&$114$ \\
                \hline
        \end{tabular}
\end{minipage}
\end{table}

\noindent  As discussed, we compare our angular momentum study of superthins and LSBs with that of a sample of six ordinary bulgeless disc galaxies and a sample of six bulgeless dwarf irregulars. The ordinary bulgeless disc galaxies are chosen from the sample of Obreschkow \& Glazebrook (2014) such that their bulge mass fraction is less than 0.05: NGC628, NGC2403, NGC2976, NGC3184, NGC3198 and NGC7793.  They have $B$ band ${\Sigma}_{s}$(0) $\sim$ 21.49 - 22.56 mag arcsec$^{-2}$ and asymptotic rotational velocities v$_{\rm{max}}$ $\sim$ 70-142 kms$^{-1}$.  \\

\subsection{Input Parameters}

GALFIT  2-D structural decompositions of the stellar surface density profiles of our sample superthins in the Spitzer 3.6 $\micron$ band were available from Salo et al (2015).
B-band photometry,  HI rotation curves and HI surface density were taken from the literature as indicated within brackets for each of our galaxies: UGC7321 (Uson \& Matthews 2003), IC5249 (O'Brien et al. 2010), IC2233 (Matthews \& Uson 2008), UGC711 (Mendelowitz et al. 2000), NGC4244 (Zschaechner et al. 2011) and FGC1540 (Kurapati et al. 2018).  
In Table 4, we present the best-fitting parameters when the rotation curves of our superthin galaxies were fitted with a Brandt profile (See \S 2.1). In Table 5, we present the best-fitting parameters when the radial HI surface density profile is fitted with a double gaussian (See \S 2.1). For the LSBs, $B$-band photometry, HI rotation curves and HI surface density profiles were all available from de Blok et al (2001). In Table 6, we present the best-fitting parameters when the rotation curves of our LSBs were fitted with a exponential profile (See \S 2.1), and in Table 7, the best-fitting parameters when the radial HI surface density profile is fitted with a double gaussian (See \S 2.1). A mass-to-light ratio constant with radius was assumed for the above studies which is reasonable for under-evolved late-type systems such as these which exhibit little or no colour variation with radius. \\

\noindent For comparison, we used the stellar angular momentum J$_s$, stellar mass M$_s$ and specific angular momentum j$_s$ as well as gas angular momentum J$_g$, gas mass M$_g$ and gas specific angular momentum j$_g$ values for the ordinary discs and dwarf irregulars as given in Obreschkow \& Glazebrook (2014) and Chowdhury \& Chengalur (2017) respectively. We may note here the J$_g$, M$_g$ and j$_g$ values for the ordinary discs included the contribution of the molecular gas. \\

\noindent 

\begin{table}
        \begin{minipage}{150mm}
	\caption{Best-fitting rotation curves: Superthins}
	\label{}
	\begin{tabular}{lccr}
		\hline
		Name & V$_{\rm{rot}}$ & R$_{\rm{max}}$ & n  \\
		     & kms$^{-1}$ & kPc &  \\
		\hline
		UGC 7321 & 103.31 $\pm{2.32}$ & 19.58 $\pm{0.74}$ & 0.983 $\pm{0.02}$ \\ \\
		IC 5249 & 108.26 $\pm{2.49}$ & 24.54 $\pm{3.87}$ & 0.835 $\pm{0.11}$ \\ \\
		IC 2233 & 109.86 $\pm{13.93}$ & 27.52 $\pm{13.34}$ & 0.678 $\pm{0.13}$ \\ \\
		UGC 711 & 104.287 $\pm{2.12}$ & 18.42 $\pm{1.45}$ & 1.83 $\pm{0.17}$ \\ \\
		NGC 4244 & 96.59 $\pm{1.24}$ & 11.69 $\pm{2.70}$ & 0.487 $\pm{0.15}$ \\ \\
		FGC 1540 & 83.36 $\pm{1.78}$ & 25.67 $\pm{6.32}$ & 0.449 $\pm{0.08}$ \\ \\
		                \hline
		                
        \end{tabular}
        \end{minipage}
\end{table}

\begin{table*}
	\centering
	\caption{Best-fitting HI surface density profiles: Superthins}
	\label{}
	\begin{tabular}{lcccccr}
		\hline
		Name & ${\Sigma}_g (0, 1)$ & $a_1$ & $r_{0,1}$ & ${\Sigma}_g (0, 2)$ & $a_2$ & $r_{0,2}$ \\
		     & M$_{\odot}$Pc$^{-2}$ & kPc & kPc & M$_{\odot}$Pc$^{-2}$ & kPc & kPc \\
		\hline
		UGC 7321 & 4.91 $\pm{0.22}$ & 8.48 $\pm{0.66}$ & 6.27 $\pm{0.26}$ & 2.51 $\pm{0.52}$ & 1.07 $\pm{0.14}$ & 3.33 $\pm{0.47}$ \\ \\
		IC 5249 & 8.05 $\pm{0.56}$ & 0.0 $\pm{0}$ & 0.25 $\pm{0.02}$ & - & - & - \\ \\
		IC 2233 & 2.24 $\pm{0.13}$ & 2.53 $\pm{0.18}$ & 1.80 $\pm{0.12}$ & 2.45 $\pm{0.14}$ & 6.15 $\pm{0.19}$ & 1.70 $\pm{0.08}$ \\ \\
		UGC 711 & 30.49 $\pm{1.5}$ & 0.0 $\pm{0}$ & 3.74 $\pm{0.22}$ & - & - & - \\ \\
		NGC 4244 & 3.93 $\pm{0.85}$ & 7.15 $\pm{0.36}$ & 1.26 $\pm{0.48}$ & 4.11 $\pm{0.43}$ & 5.01 $\pm{0.34}$ & 4.61 $\pm{0.22}$ \\ \\
		FGC 1540 & 4.09 $\pm{0.30}$ & 2.48 $\pm{0.87}$ & 5.73 $\pm{0.83}$ & 1.30 $\pm{0.39}$ & 5.08 $\pm{0.10}$ & 1.20 $\pm{0.26}$ \\ \\
		\hline
	\end{tabular}
\end{table*}

\begin{table}
\begin{minipage}{150 mm}
	\caption{Best-fitting rotation curves: LSBs}
	\label{}
	\begin{tabular}{lcr}
		\hline
		Name & $\alpha$ & $\beta$ \\
		     & kms$^{-1}$ & kPc \\
		\hline
		F563-V2 & 118.13 $\pm{2.7}$  & 1.9 $\pm{0.14}$ \\ \\ 
		F574-1 & 99.10  $\pm{0.79}$ & 2.62 $\pm{0.08}$ \\ \\
		F583-1 & 88.84 $\pm{1.04}$ & 3.29 $\pm{0.09}$ \\ \\
		F583-4 & 66.19 $\pm{2.16}$ & 1.61 $\pm{0.21}$ \\ \\
		F568-V1 & 117.32 $\pm{1.10}$ & 2.19 $\pm{0.08}$ \\ \\
		F563-1 & 109.59 $\pm{2.56}$ & 2.62 $\pm{0.30}$ \\ \\
		F579-V1 & 110.65 $\pm{1.28}$ & 1.05 $\pm{0.07}$ \\ \\
		F568-1 & 139.23 $\pm{2.26}$ & 2.29 $\pm{0.14}$ \\ \\
		F568-3 & 107.01 $\pm{4.99}$ & 3.54 $\pm{0.40}$ \\ \\
		\hline
	\end{tabular}
	\end{minipage}
\end{table}

\begin{table*}
	\centering
	\caption{Best-fitting HI surface density profiles: LSBs}
	\label{}
	\begin{tabular}{lcccccr}
		\hline
		Name & ${\Sigma}_g (0, 1)$ & $a_1$ & $r_{0,1}$ & ${\Sigma}_g (0, 2)$ & $a_2$ & $r_{0,2}$ \\
		     & M$_{\odot}$Pc$^{-2}$ & kPc & kPc & M$_{\odot}$Pc$^{-2}$ & kPc & kPc \\
		\hline
		F563-V2 & 6.89 $\pm{0.11}$ & 4.57 $\pm{0.11}$ & 2.53 $\pm{0.07}$ & 5.21 $\pm{0.25}$ &0.0 $\pm{0.09}$ & 1.74 $\pm{0.15}$ \\ \\
		F574-1 & 2.31 $\pm{0.08}$ & 0.51 $\pm{0.98}$ & 6.18 $\pm{0.32}$ & 1.60 $\pm{0.24}$ & 7.70 $\pm{0.18}$ & 3.17 $\pm{0.22}$ \\ \\
		F583-1 & 0.35 $\pm{0.12}$ & 4.66 $\pm{0.24}$ & 0.96 $\pm{0.37}$ & 4.37 $\pm{0.11}$ & 3.94 $\pm{0.08}$ & 4.14 $\pm{0.11}$ \\ \\
		F583-4 & 1.09 $\pm{0.28}$ & -0.27 $\pm{0.33}$ & 1.49 $\pm{0.37}$ & 1.41 $\pm{0.10}$ & 3.23 $\pm{0.44}$ & 2.84 $\pm{0.22}$ \\ \\
		F568-V1 & 1.76 $\pm{0.32}$ & 7.90 $\pm{0.12}$ & 2.28 $\pm{0.21}$ & 3.56 $\pm{0.14}$ & 2.97 $\pm{0.37}$ & 4.36 $\pm{0.22}$ \\ \\
		F563-1 & 4.61 $\pm{0.45}$ & -3.74 $\pm{3.20}$ & 11.40 $\pm{1.70}$ & 0.76 $\pm{0.24}$ & 4.75 $\pm{0.49}$ & 2.06 $\pm{0.70}$ \\ \\
		F579-V1 & 3.56 $\pm{0.05}$ & 4.11 $\pm{0.13}$ & 5.14 $\pm{0.14}$ & - & - & - \\ \\
		F568-1 & 7.94 $\pm{0.15}$ & 4.41 $\pm{0.01}$ & 4.06 $\pm{0.11}$ & - & - & - \\ \\
		F568-3 & 4.63 $\pm{0.13}$ & 3.95 $\pm{0.23}$ & 4.98 $\pm{0.25}$ & - & - & - \\ \\
		\hline
	\end{tabular}
\end{table*}

\section{Results} 
In Table 8, we present our calculated values of the mass M$_{\rm{stars}}$, the angular momentum J$_{\rm{stars}}$ and the specific angular momentum j$_{\rm{stars}}$ of our sample galaxies in the 3.6 $\micron$ band. Our results indicate that IC5249 has the highest value of j$_{\rm{stars}}$ among our superthins followed by FGC1540, UGC7321, UGC711, NGC4244 and IC2233. j$_{\rm{stars}}$ values for our sample superthins range between 10$^{5.44}$ - 10$^{5.95}$ comparable to 10$^{5.03}$ - 10$^{6.03}$ (both in j/kms$^{-1}$ pc units) found for a sample of 5 bulgeless ordinary spirals in Obreschkow \& Glazebrook (2014). We stress here that the 3.6$\micron$ band traces the old stellar population which also constitutes the main mass component of the galactic stellar disc. However, the attribute \emph{superthin} for our sample galaxies originated from their razor-thin appearance in the optical or B-band, and they may not appear to be be quite \emph{superthin} in the 3.6$\micron$ band. However, the optical or the $B$-band, which traces the young stellar population, may be highly obscured by dust and therefore may not be the ideal choice to estimate the mass or angular momentum of stellar discs. However, this effect is expected to be less severe in case of low surface brightness galaxies like the superthins due to their low dust content (de Blok et al. 2001). We therefore repeat the above study using structural decompositions in the $B$-band for all our sample galaxies, barring FGC1540 for which no optical data were available. In Table 9, we summarize the M$_{\rm{stars}}$, J$_{\rm{stars}}$ and j$_{\rm{stars}}$ values of our sample galaxies in the $B-$ band. We find that the specific angular momenta of our superthin stellar discs in the $B$-band are less than a factor of 2 higher than those in the 3.6 $\micron$ band, except for IC2233 which has comparable j$_{\rm{stars}}$ values in both the bands. Besides our sample galaxies follow the same trend in the magnitude of their j$_{\rm{stars}}$ values as in the 3.6 $\micron$ band. In Table 10, we present our calculated values for our LSB sample in the $B$ band. j$_{\rm{stars}}$ values for our sample LSBs range between 10$^{5.3}$ - 10$^{6.2}$ and is comparable to those of the superthins. \\

\noindent The calculated stellar mass M$_s$ varies between 10$^{8.3}$ - 10$^{9.4}$ M$\odot$, 10$^{8.5}$ - 10$^{9.9}$ M$\odot$ and 10$^{9.5}$ - 10$^{10.3}$ M$\odot$ for the superthins, LSBs, and ordinary discs respectively.
All the mass estimates correspond to the 3.6$\mu$ band except for the LSBs for which the calculated stellar masses correspond to the $B$-band.
Therefore, our superthins and LSBs have M$_s$ about an order of magnitude higher than those of the ordinary discs. Further, in the $B$-band, M$_s$ for our sample superthins are found to be a few times higher than those in the 3.6$mu$ band. \\

\begin{table*}
	\centering
	\caption{Angular momenta of the stellar discs of the superthins: 3.6$\mu$ band}
	\label{tab:example_table}
	\begin{threeparttable}
	\begin{tabular}{lccccccccr} 
		\hline
		Name & &M$_{\rm{stars}}$\tnote{1} & & &J$_{\rm{stars}}$\tnote{2} & & & j$_{\rm{stars}}$\tnote{3} & \\
		\hline
                     & Disc 1 & Disc 2 & Total & Disc 1 & Disc 2 & Total&  Disc 1& Disc 2& Total \\
		\hline
		UGC7321 &9.10 &9.05 &9.38 &15.09$^{+0.01}_{-0.01}$ &14.58$^{+0.02}_{-0.02}$ &15.21$^{+0.01}_{-0.01}$ &6.00$^{+0.01}_{-0.01}$ &5.53$^{+0.02}_{-0.02}$ &5.83$^{+0.01}_{-0.01}$\\ \\
		IC5249  &8.97&8.18&9.04&14.98$^{+0.04}_{-0.05}$ &13.34$^{+0.06}_{-0.07}$ &15.00$^{+0.04}_{-0.05}$ &6.01$^{+0.04}_{-0.05}$ &5.16$^{+0.06}_{-0.07}$ &5.95$^{+0.04}_{-0.05}$\\ \\
		IC2233 &8.21&7.70&8.33&13.75$^{+0.14}_{-0.19}$&12.61$^{+0.15}_{-0.24}$&13.78$^{+0.13}_{-0.18}$ &5.53$^{+0.14}_{0.19}$ &4.91$^{+0.15}_{-0.24}$ &5.45$^{+0.13}_{-0.18}$\\\\
		UGC711  &8.62&&8.62&14.07$^{+0.03}_{-0.04}$&&14.07$^{+0.03}_{-0.04}$ &5.45$^{+0.03}_{-0.04}$ & &5.45$^{+0.03}_{-0.04}$\\ \\
		NGC4244  &8.57&&8.57&14.02$^{+0.06}_{-0.08}$&&14.02$^{+0.06}_{-0.08}$ & 5.44$^{+0.06}_{-0.08}$ & &5.44$^{+0.06}_{-0.08}$\\ \\
		FGC1540 &8.38&7.64&8.46&14.09$^{+0.05}_{-0.06}$&12.68$^{+0.04}_{-0.04}$&14.11$^{+0.05}_{-0.05}$ &5.71$^{+0.05}_{-0.06}$ &5.03$^{+0.04}_{-0.04}$ &5.65$^{+0.05}_{-0.05}$\\ \\
		
		\hline
	\end{tabular}
	\begin{tablenotes}\footnotesize
		\item[1] in Log(M/M$_{\odot}$) 
		\item[2] in Log(J/(M$_{\odot}$kms$^{-1}$ pc))
		\item[3] in Log(j/(kms$^{-1}$ pc)
	\end{tablenotes}
	\end{threeparttable}
\end{table*}

\begin{table}
	\centering
	\caption{Angular momenta of the stellar discs of the superthins: $B$-band}
	\label{tab:example_table}
	\begin{threeparttable}
	\begin{tabular}{lccr} 
		\hline
		Name &M$_{\rm{stars}}$\tnote{1} & J$_{\rm{stars}}$\tnote{2} & j$_{\rm{stars}}$\tnote{3}  \\
		\hline
		UGC7321 & 9.67 & 15.60$^{+0.01}_{-0.02}$ & 5.94$^{+0.01}_{-0.02}$\\ \\
		IC5249     & 9.34 & 15.50$^{+0.04}_{-0.04}$ &  6.15$^{+0.04}_{-0.04}$\\ \\
		IC2233     & 8.74 & 14.05$^{+0.14}_{-0.21}$  & 5.31$^{+0.14}_{-0.21}$ \\ \\
		UGC711   & 9.31 & 15.02$^{+0.03}_{-0.03}$ &  5.70$^{+0.03}_{-0.03}$\\ \\
		NGC4244 & 9.61 & 14.71$^{+0.06}_{-0.06}$ &  5.55$^{+0.06}_{-0.06}$\\ \\
		\hline
	\end{tabular}
	\begin{tablenotes}\footnotesize
		\item[1] in Log(M/M$_{\odot}$)
		\item[2] in Log(J/(M$_{\odot}$kms$^{-1}$ pc))
		\item[3] in Log(j/(kms$^{-1}$ pc)
	\end{tablenotes}
	\end{threeparttable}
\end{table}

\begin{table}
	\centering
	\caption{Angular momenta of the stellar discs of the LSBs: $B$-band}
	\label{tab:example_table}
	\begin{threeparttable}
	\begin{tabular}{lccr} 
		\hline
		Name &M$_{\rm{stars}}$\tnote{1} & J$_{\rm{stars}}$\tnote{2} & j$_{\rm{stars}}$\tnote{3}  \\
		\hline
		F563-V2 & 9.46 &15.10 $\pm$ 0.01 & 2.65 $\pm$ 0.01 \\ \\
		F574-1 & 9.60 & 15.51 $\pm$ 0.00 & 2.91 $\pm$ 0.00 \\ \\
		F583-1 & 8.45 &13.74  $\pm$ 0.01 & 2.29 $\pm$ 0.01 \\ \\
		F583-IV & 9.01 &14.54 $\pm$ 0.02 & 2.53 $\pm$ 0.02 \\ \\
		F563-1 & 9.11 & 14.84 $\pm$ 0.01 & 2.74 $\pm$ 0.01 \\ \\
		F568-V1 & 9.34 &15.19 $\pm$ 0.00 & 2.84 $\pm$ 0.00 \\ \\
		F568-1 & 9.58 & 15.74 $\pm$ 0.01 & 3.16 $\pm$ 0.01 \\ \\
		F568-3 & 9.62 & 15.50 $\pm$ 0.02 & 2.88 $\pm$ 0.02 \\ \\
		F579-V1 & 9.95 & 16.00 $\pm$ 0.01 & 3.05 $\pm$ 0.01 \\ \\
		\hline
	\end{tabular}
	\begin{tablenotes}\footnotesize
		\item[1] in Log(M/M$_{\odot}$)
		\item[2] in Log(J/(M$_{\odot}$kms$^{-1}$ pc))
		\item[3] in Log(j/(kms$^{-1}$ pc)
	\end{tablenotes}
	\end{threeparttable}
\end{table}

 We next study if a higher value of the stellar specific angular momentum j$_{\rm{stars}}$ leads to a larger disc size irrespective of the depth of the potential well as indicated by $V_{\rm{rot}}$. In Figure 1, we present the regression line fitted to the 1.68 $R_D$ versus j$_{\rm{s}}$ values of the ordinary bulgeless disc galaxies along with its 95.4$\%$ confidence interval, and superpose the data for the LSB (solid squares) and superthin (solid triangles) samples on it. 1.68 $R_D$ could be taken to be the half-stellar mass radius of the galaxy (Jiang et al. 2018). 
 We find that four out of the six superthins and most of the LSBs have systematically larger $R_D$ values compared to ordinary bulgeless disc galaxies for a given value of j$_{\rm{s}}$. 
 
\begin{figure}
\centering
\includegraphics[width=.5\textwidth]{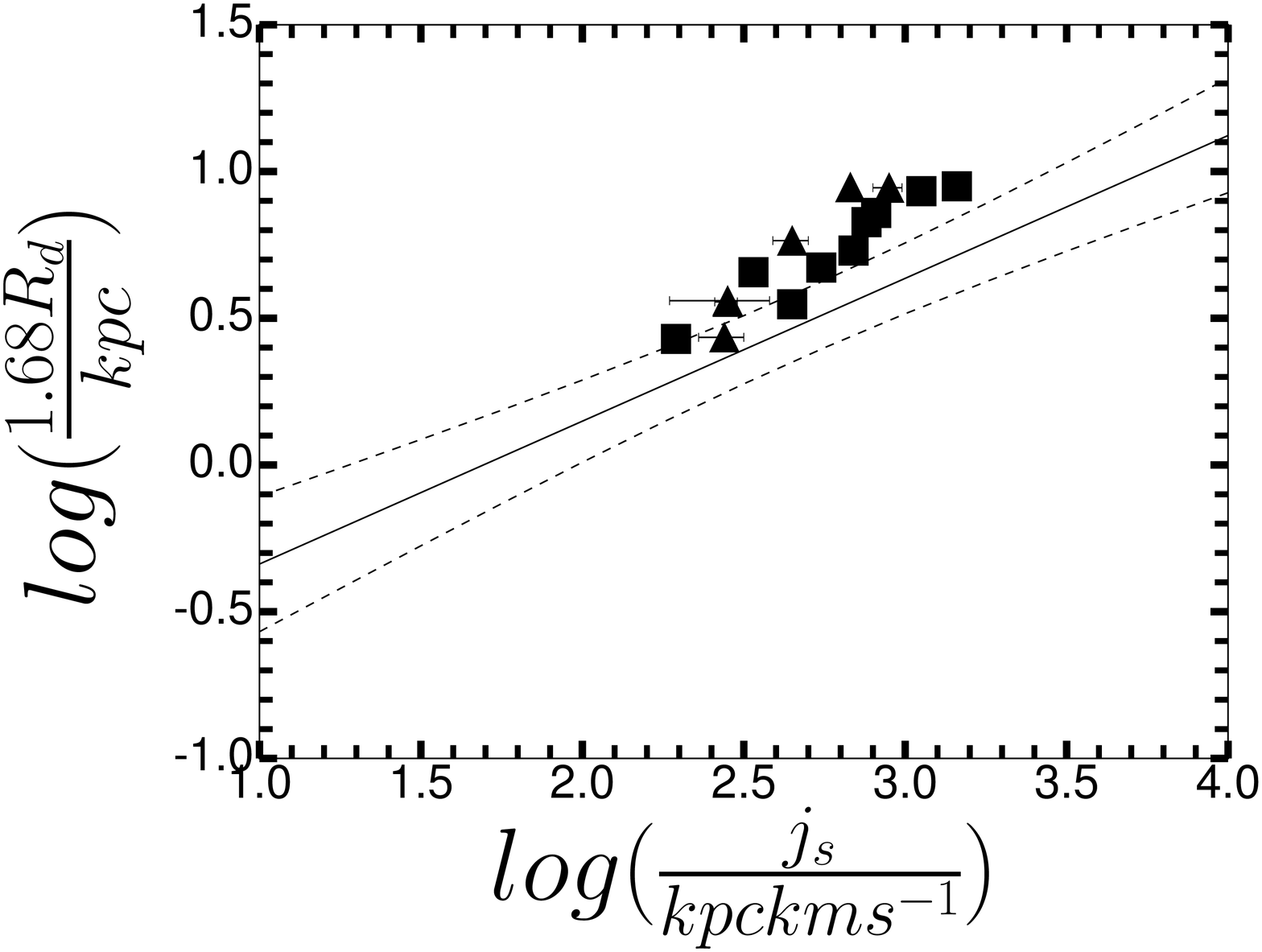}
\caption{Plot of the characteristic radius of an exponential stellar disc $\sim$ 1.68 $R_D$, $R_D$ being the exponential stellar disc scale length versus specific angular momentum of the stellar disc j$_{\rm{s}}$. The solid line represents the regression line fit to the ordinary bulgeless spirals and the dotted line the 95.4$\%$ confidence interval of the same.The slope and intercept of the line is given by $0.49\pm{0.103}$ and $-0.82\pm{0.277}$ respectively.  Superposed on the plot are the data for the superthins (solid triangles), LSBs (solid squares). The stellar photometry corresponds to the 3.6 $\mu$ band for all the galaxy types except for the LSBs for which it is the B-band. Also, for galaxies 
with two exponential stellar discs, $R_D$ corresponds to the disc scale length of the larger disc.}
\label{fig:nfw_5249}
\end{figure} 
 
\noindent As discussed earlier, the size of the stellar disc is primarily regulated by a balance between its the specific angular momenta and its radial gravitational field; a stellar disc with a given specific angular momentum is expected to have a larger disc size in a shallower gravitational potential well and vice-versa. In Figure 2, we fit a regression line to the $j_{\rm{s}}$ versus $V_{\rm{rot}}$ data for the ordinary spirals along with the 95.4$\%$ confidence band, $V_{\rm{rot}}$ being the asymptotic rotational velocity. We also superpose the $j_{\rm{s}}$ versus $v_{\rm{rot}}$ data for the superthins (solid triangles) and LSBs (solid squares) on this plot. We find that for three out of our six superthins  and seven out of our nine LSBs, the $j_{\rm{s}}$ values lie above the 95.4 $\%$ confidence band of the $j_{\rm{s}}$ - $V_{\rm{rot}}$ regression line for ordinary bulgeless disc galaxies. We note that although the rest of the data points for the superthins lie within the 95.4$\%$ confidence band for the ordinary spirals, superthins and LSBs have systematically higher values of $j_{\rm{s}}$  momenta as compared to the ordinary spirals for a given value of the asymptotic rotational velocity $v_{\rm{rot}}$.

\begin{figure}

\centering
\includegraphics[width=.5\textwidth]{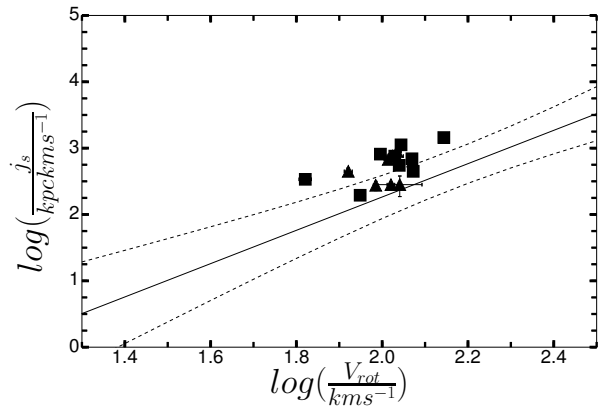}
\caption{Plot of the specific angular momenta of the stellar discs  j$_{\rm{s}}$ versus the asymptotic rotational velocities V$_{\rm{rot}}$. The solid line represents the regression line fit to the ordinary bulgeless spirals and the dotted line the 95.4$\%$ confidence interval of the same.The slope and intercept of the line is given by $2.51\pm{0.72}$ and $-2.76\pm{1.55}$ respectively. Superposed on the plot are the $j_{\rm{s}}$ versus $V_{\rm{rot}}$ data for the superthins (filled triangles), LSBs (filled squares)}
\label{fig:nfw_5249}
\end{figure}

\noindent However, $V_{\rm{rot}}$ indicates  the depth of the gravitational potential well of the galaxy. However, the actual radial gravitational field and hence the rotational velocity $V_{\rm{rot}}$ at any intermediate galacto-centric radius $R$ depends on the slope of the rotation curve. For a steeply rising rotation curve, the mean rotational velocity for the disc is close to $V_{\rm{rot}}$ whereas for a slowly rising rotation curve, the mean rotational velocity may be quite smaller than $V_{\rm{rot}}$. In fact,
we checked that even within each of the different galaxy types like the superthins, LSBs and the ordinary discs, there are steeply-rising to slowly-rising rotation curves.
Among our superthins, for example, UGC7321 has the most steeply-rising rotation curve, followed by IC5249, NGC4244, FGC1540, IC2233 and UGC 711.
In order to determine how $V_{\rm{rot}}$ compares with the mean rotational velocity, in Figure 3, we present a linear fit to the 2 R$_D$ $V_{\rm{rot}}$ versus $j_{\rm{stars}}$ data for all different galaxy samples taken together: LSBs, superthins and ordinary disc galaxies; 2 R$_D$ $v_{\rm{rot}}$ is representative of the stellar specific angular momentum according to Mo, Mao \& White (1998). Interestingly, the slope and intercept of the regression line fit for all the galaxies taken together are 0.97 $\pm$ 0.03 and 72.78 $\pm$ 23.33, respectively. This plot and the fit parameters together indicate that the mean rotational velocity is roughly the same for all galaxy types, except for a couple of superthins, which seem to have a smaller mean velocity and hence a larger disc size for a given value of $j_{\rm{s}}$. Therefore, we may conclude that some superthins and LSBs may have higher values of  $j_{\rm{stars}}$ for a given values of $V_{\rm{rot}}$ and hence larger disc sizes compared to an ordinary disc galaxy. This, in turn, may lead to a larger value of the planar-to-vertical axes ratio as compared  an ordinary bulgeless spiral with the same value of $v_{\rm{rot}}$.Therefore, a high value of the specific angular momentum may possibly drive the existence of superthin stellar discs in some LSBs.
\begin{figure}
\centering
\includegraphics[width=.5\textwidth]{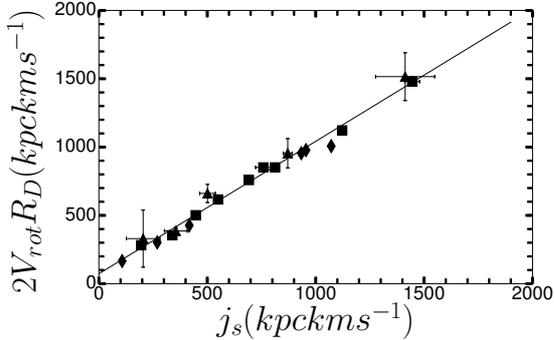}
\caption{Plot of 2$V_{\rm{rot}}$$R_D$ versus the stellar specific angular momentum j$_{\rm{s}}$ for superthins(triangles), LSBs (squares) , and ordinary discs (diamonds). The stellar photometry corresponds to that of the $B$-band.}
\label{fig:nfw_5249}
\end{figure}


We next study the specific angular momenta of the gas discs to check for correlation, if any, between the specific angular momenta of the stellar and the gas discs in our sample superthins as the gas discs constitutes the progenitors of the stellar discs.
In Table 11, we present the mass M$_{\rm{g}}$, the angular momentum J$_{\rm{g}}$ and the specific angular momentum j$_{\rm{g}}$ of our 
sample superthin galaxies as obtained from HI 21cm radio-synthesis observations, including corrections for the presence of Helium, and In Table 12, we present our corresponding values for the LSB sample. The gas mass M$_g$ varies between 10$^{6.5}$ - 10$^{10.0}$ M$\odot$, 10$^{8.4}$ - 10$^{9.4}$ M$\odot$ for the superthins and the LSBs respepctively. In comparison, it varies between 10$^{8.4}$ - 10$^{10.1}$ M$\odot$ for the ordinary discs. Besides,  the respective median M$_g$ values for the superthins, LSBs and ordinary discs are 10$^{9.2}$ M$\odot$, 10$^{9.1}$ M$\odot$ and 10$^{9.5}$ M$\odot$. Therefore the superthins and LSBs have M$_g$ values comparable with those of the ordinary discs. We further note  j$_{\rm{gas}}$ is in general higher than j$_{\rm{s}}$ for all our sample superthins. In fact,this is the reflection of the fact that the gas disc larger in size than that of the stars. Following the stellar case, in Figure 4, we present a regression line fit to the $j_{\rm{gas}}$ versus $V_{\rm{rot}}$ data for the ordinary bulgeless spirals (black line) using the data of Obreschkow \& Glazebrook (2014) along with its 95.4$\%$ confidence band (short-dashed line). We also superpose the $j_{\rm{g}}$ versus $V_{\rm{rot}}$ data for the superthins (filled triangles) and LSBs (solid spheres) on this plot. Interestingly, we note that unlike the stellar component, superthins are consistent with the $j_{\rm{g}}$  -  $V_{\rm{rot}}$ relation of the ordinary spirals. However, we may note here that several of our sample galaxies have HI holes which indicates that the low angular momentum gas has been removed from them by supernovae explosions or other feedback effects. Therefore our calculated values of $j_{\rm{g}}$ may not be representative of the primordial or original$j_{\rm{gas}}$ of these galaxies, and hence it is not trivial to link the  $j_{\rm{g}}$ to $j_{\rm{s}}$ without careful modelling of feedback effects.

\begin{table*}
	\centering
	\caption{Angular momenta of the gas discs of the superthins}
	\label{tab:example_table}
	\begin{threeparttable}
	\begin{tabular}{lccccccccr} 
		\hline
		Name & &M$_{\rm{gas}}$\tnote{1} & & &J$_{\rm{gas}}$ \tnote{2} & & & j$_{\rm{gas}}$\tnote{3} & \\
                     & Disc 1 & Disc 2 & Total & Disc 1 & Disc 2 & Total&  Disc 1& Disc 2& Total \\
		\hline
	UGC7321 & 9.63$^{+0.03}_{-0.03}$ & 8.40$^{+0.08}_{-0.10}$ & 9.65$^{+0.03}_{-0.03}$ & 15.72$^{+0.}_{-0.}$ &13.95$^{+0.09}_{-0.11}$ & 15.73$^{+0.01}_{-0.01}$ & 6.10$^{+0.03}_{-0.03}$ & 5.55$^{+0.12}_{-0.16}$ & 6.10$^{+0.03}_{-0.03}$ \\ \\
        IC5249 & 6.51$^{+0.03}_{-0.03}$ & & 6.51$^{+0.03}_{-0.03}$ & 10.06$^{+0.13}_{-0.19}$ & & 10.36$^{+0.13}_{-0.19}$ & 3.55$^{+0.13}_{-0.19}$ & & 3.55$^{+0.13}_{-0.19}$ \\ \\
        IC2233 & 8.21$^{+0.04}_{-0.04}$ & 8.61$^{+0.03}_{-0.03}$ & 8.75$^{+0.03}_{-0.03}$& 13.62$^{+0.12}_{-0.17}$ & 14.36$^{+0.10}_{-0.12}$ & 14.44$^{+0.08}_{-0.10}$ & 5.41$^{+0.13}_{-0.18}$ & 5.76$^{+0.10}_{-0.13}$ & 5.68$^{+0.09}_{-0.11}$ \\ \\ 
        UGC711 & 10.02$^{+0.05}_{-0.06}$ & & 10.02$^{+0.05}_{-0.06}$ & 15.94$^{+0.02}_{-0.02}$ & & 15.94$^{+0.02}_{-0.02}$  &5.92$^{+0.05}_{-0.06}$ & & 5.92$^{+0.05}_{-0.06}$\\ \\
        NGC4244 & 8.75$^{+0.16}_{-0.25}$ & 9.20$^{+0.05}_{-0.05}$ & 9.33$^{+0.06}_{-0.07}$& 14.59$^{+0.03}_{-0.03}$&15.11$^{+0.07}_{-0.07}$ & 15.22$^{0.05}_{-0.06}$ & 5.85$^{+0.03}_{-0.03}$ & 5.91$^{+0.07}_{-0.07}$ & 5.89$^{+0.08}_{-0.08}$ \\ \\ 
        FGC1540	& 9.14$^{+0.14}_{-0.20}$ & 8.10$^{+0.09}_{-0.20}$ & 9.18$^{+0.07}_{-0.11}$ & 14.95$^{+0.16}_{-0.25}$ & 13.66$^{+0.15}_{-0.23}$ & 14.97$^{+0.15}_{-0.23}$ & 5.81$^{+0.17}_{-0.30}$ & 5.57$^{+0.19}_{-0.30}$ & 5.80$^{+0.17}_{-0.27}$ \\ \\
		\hline
	\end{tabular}
	\begin{tablenotes}\footnotesize
		\item[1] In Log(M/M$_{\odot}$)
		\item[2] In Log(J/(M$_{\odot}$kms$^{-1}$ pc))
		\item[3] In Log(j/(kms$^{-1}$ pc))
	\end{tablenotes}
	\end{threeparttable}
\end{table*}

\begin{table}
	\centering
	\caption{Angular momenta of the gas discs of the LSBs}
	\label{tab:example_table}
	\begin{threeparttable}
	\begin{tabular}{lccr} 
		\hline
		Name &M$_{\rm{g}}$\tnote{1}  & J$_{\rm{g}}$\tnote{2}  & j$_{\rm{g}}$\tnote{3}  \\
		\hline
		F563-V2 & 9.13 $\pm$ 0.02  &14.93 & 2.80 $\pm$ 0.03 \\ \\
		F574-1   &  9.10 $\pm$ 0.05  & 15.0 & 2.90$^{-0.11}_{+0.08}$\\ \\
		F583-1  &  9.10 $\pm$ 0.02 & 14.85 & 2.75 $\pm$ 0.03 \\ \\
		F583-IV & 8.36$^{-0.07}_{+0.06}$ &13.87 & 2.51$^{-0.14}_{+0.11}$ \\ \\
		F563-1 & 9.40$^{+0.18}_{-0.33}$ & 15.53$^{+0.24}_{-0.58}$ & 3.12$^{+0.28}_{-1.0}$ \\ \\
		F568-V1 & 9.14$^{+0.04}_{-0.05}$ & 15.07$^{0.05}_{-0.06}$ & 2.93$^{+0.07}_{-0.08}$ \\ \\
		F568-1 & 9.37 $\pm$ 0.02 & 15.37 $\pm$ 0.03 & 2.99 $\pm$ 0.03 \\ \\
		F568-3 & 9.22 $^{+0.02}_{-0.03}$ & 15.12 $\pm$ 0.06 & 2.90 $^{+0.06}_{-0.08}$ \\ \\
		F579-V1 & 9.13 $\pm$ 0.02 & 15.11 $\pm$ 0.03 & 2.97 $^{+0.03}_{-0.04}$ \\ \\
		\hline
	\end{tabular}
	\begin{tablenotes}\footnotesize
		\item[1] In Log(M/M$_{\odot}$)
		\item[2] In Log(J/(M$_{\odot}$kms$^{-1}$ pc))
		\item[3] In Log(j/(kms$^{-1}$ pc))
	\end{tablenotes}
	\end{threeparttable}
\end{table}

\begin{figure}

\centering
\includegraphics[width=.5\textwidth]{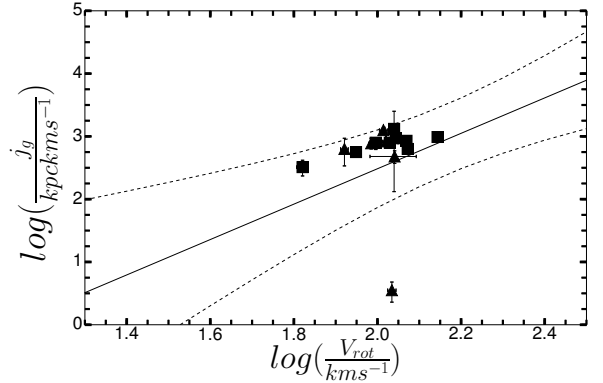}

\caption{Plot of the specific angular momenta of the gas discs  j$_{\rm{g}}$ versus the asymptotic rotational velocities V$_{\rm{rot}}$. The solid line represents the regression line fit to the ordinary bulgeless spirals and the dotted line the 95.4$\%$ confidence interval of the same.The slope and intercept of the line is given by $2.82\pm{1.36}$ and $-3.15\pm{2.95}$ respectively . Superposed on the plot are the $j_{\rm{g}}$ versus $V_{\rm{rot}}$ data for the superthins (solid triangles), LSBs (solid squares).}
\end{figure}

Finally, in Tables 13 and 14, we present the total baryonic mass M$_{\rm{b}}$, the baryonic angular momentum J$_{\rm{b}}$ and the baryonic specific angular momentum 
j$_{\rm{b}}$ of our sample superthin galaxies and LSBs respectively. In case of the superthins, both j$_{\rm{b}}$ and M$_{\rm{b}}$ closely reflect
 j$_{\rm{g}}$ and M$_{\rm{g}}$ respectively.  In contrast, the j$_{\rm{b}}$ and M$_{\rm{b}}$ for LSBs and ordinary spirals are not indicative of the values of any single disc component i.e., stars or gas.  In fact, for ordinary spirals, M$_{\rm{b}}$ is significantly higher than M$_{\rm{g}}$, and j$_{\rm{b}}$ is
 quite less than j$_{\rm{g}}$ (Obreschkow \& Glazebrook 2014). In Figure 5, we present a regression line fit to the $j_{\rm{b}}$ versus $V_{\rm{rot}}$ data for the ordinary bulgeless spirals using the data of Obreschkow \& Glazebrook (2014) along with its 95.4$\%$ confidence band. We also superpose the $j_{\rm{b}}$ versus $V_{\rm{rot}}$ data for the superthins (solid triangles)and LSBs (solid squares) on this plot. As in the gas disc case, for a given value of V$_{\rm{rot}}$,
 all of superthins and LSBs have j$_{\rm{b}}$ comparable with the ordinary discs. 
 (See Appendix for a similar study involving the correlation of the 
j$_{\rm{s}}$ - M$_{\rm{s}}$, j$_{\rm{g}}$ - M$_{\rm{g}}$ and 
j$_{\rm{b}}$ - M$_{\rm{b}}$ values i.e., the Fall Relation.) \\

\begin{table}
	\centering
	\caption{Total angular momenta of the baryonic discs of the superthins}
	\label{tab:example_table}
	\begin{threeparttable}
	\begin{tabular}{lccr} 
		\hline
		Name & M$_{\rm{baryons}}$\tnote{1}& J$_{\rm{baryons}}$\tnote{2} & j$_{\rm{baryons}}$\tnote{3} \\
                     
		\hline
UGC7321 & 9.84$^{+0.02}_{-0.02}$ &15.84$^{0.00}_{-0.00}$ & 6.01$^{+0.02}_{-0.02}$ \\
IC5249 & 9.04$^{+0.00}_{-0.00}$ & 15.00$^{+0.04}_{-0.05}$ &5.95$^{+0.04}_{-0.05}$ \\
IC2233& 8.89$^{+0.02}_{-0.02}$ & 14.52$^{+0.07}_{-0.09}$  &5.63$^{+0.08}_{-0.09}$ \\
UGC711& 10.04$^{+0.05}_{-0.06}$ & 15.95$^{+0.02}_{-0.02}$ &5.91$^{+0.06}_{-0.06}$ \\
NGC4244& 9.40$^{+0.05}_{-0.06}$ & 15.25$^{+0.05}_{-0.05}$ &5.85$^{+0.07}_{-0.08}$ \\
FGC1540 & 9.25$^{+0.07}_{-0.09}$ & 15.03$^{+0.13}_{-0.20}$ &5.77$^{+0.15}_{-0.23}$ \\
		\hline
	\end{tabular}
	\begin{tablenotes}\footnotesize
		\item[1] In Log(M/M$_{\odot}$)
		\item[2] In Log(J/(M$_{\odot}$kms$^{-1}$ pc))
		\item[3] In Log(j/(kms$^{-1}$ pc))
	\end{tablenotes}
	\end{threeparttable}
\end{table}

\begin{table}
	\centering
	\caption{Angular momenta of the baryonic discs of the LSBs}
	\label{tab:example_table}
	\begin{threeparttable}
	\begin{tabular}{lccr} 
		\hline
		Name &M$_{\rm{stars}}$\tnote{1}&J$_{\rm{stars}}$\tnote{2} & j$_{\rm{stars}}$\tnote{3}  \\
		\hline
		F563-V2 &  9.62 $\pm$ 0.01& 15.32 &2.70 $\pm$ 0.01 \\ \\
		F574-1   &  9.72 $\pm$ 0.01 & 15.63 &2.91 $\pm$ 0.02\\ \\
		F583-1  &   9.19 $\pm$ 0.02 & 14.88 &2.69$^{-0.03}_{+0.02}$\\ \\
		F583-IV & 9.10 $\pm$ 0.01 &14.62 & 2.52 $\pm$ 0.02\\ \\
		F563-1 &9.58 $^{+0.13}_{-0.19}$ & 15.61$^{+0.21}_{-0.41}$ & 3.03$^{+0.23}_{-0.53}$ \\ \\
		F568-V1 & 9.55 $\pm$ 0.02 & 15.43 $\pm$ 0.02 & 2.88 $\pm$ 0.03 \\ \\
		F568-1 & 9.79 $\pm$ 0.01 & 15.89 $\pm$ 0.01 & 3.10 $\pm$ 0.01 \\ \\
		F568-3 & 9.76 $\pm$ 0.01 & 15.65 $\pm$ 0.02 & 2.88$^{+0.02}_{-0.03}$ \\ \\
		F579-V1 &10.01 $\pm$ 0.002 & 16.05 $\pm$ 0.005 & 3.04 $\pm$ 0.01 \\ \\
		\hline
	\end{tabular}
	\begin{tablenotes}\footnotesize
		\item[1] In Log(M/M$_{\odot}$)
		\item[2] In Log(J/(M$_{\odot}$kms$^{-1}$ pc))
		\item[3] In Log(j/(kms$^{-1}$ pc))
	\end{tablenotes}
	\end{threeparttable}
\end{table}

\begin{figure}

\centering
\includegraphics[width=.5\textwidth]{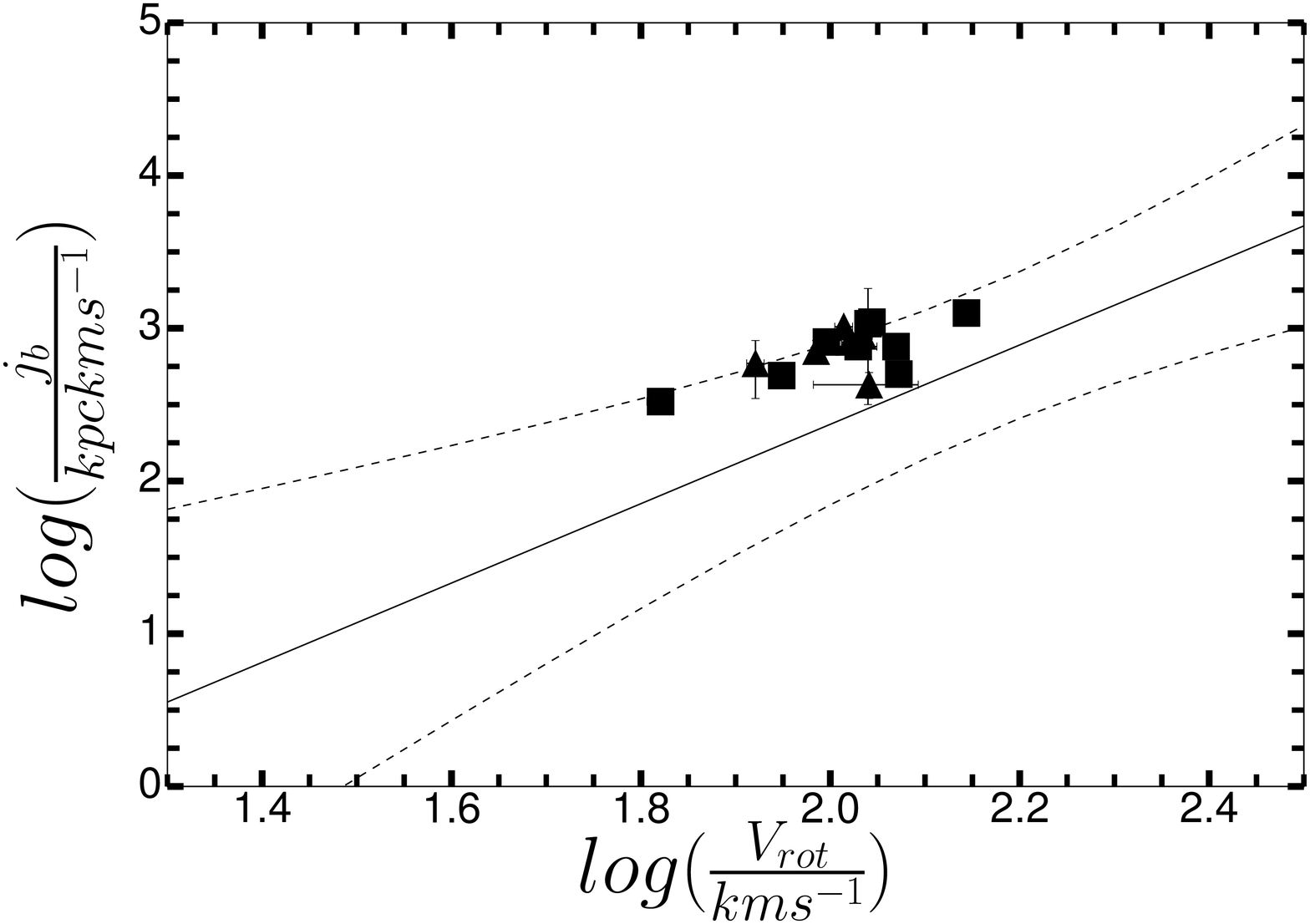}

\caption{Plot of the specific angular momenta of the baryonic discs  j$_{\rm{b}}$ versus the asymptotic rotational velocities V$_{\rm{rot}}$. The solid line represents the regression line fit to the ordinary bulgeless spirals and the dotted line the 95.4$\%$ confidence interval of the same.The slope and intercept of the line is given by $2.60\pm{1.16}$ and $-2.83\pm{2.52}$ respectively.  Superposed on the plot are the $j_{\rm{b}}$ versus $V_{\rm{rot}}$ data for the superthins (solid triangles), LSBs (soild squares). }
\label{fig:nfw_5249}
\end{figure}

\textbf{Size-Mass Relation:} We also study the size-mass relation in our superthin and LSB samples following earlier studies in the literature, which showed the afore-mentioned scaling relation strongly depends on the galaxy morphology, with late-types having characteristically larger disc sizes compared to early-type galaxies (See, for example, Shen et al. 2003). In Figure 6, we present the regression line fit to 1.68 R$_D$, which is the half stellar mass radius and hence a proxy for the stellar disc size versus the stellar mass M$_{\rm{s}}$ to the data for the ordinary spirals. Superposed on it are the data points  for the superthins (solid triangles), LSBs (solid squares). We observe that the superthins and LSBs all lie outside the 95.4$\%$ band of the ordinary spirals, implying that they have larger disc size for a given stellar mass as compared to the ordinary spirals. We observe that all the LSBs and the superthins lie outside the 95.4$\%$ confidence band of the of the ordinary spirals. This indicates that they have significantly larger disc size compared to ordinary spirals with the same stellar mass, which is in line with the $j_{\rm{s}}$ - $M_{\rm{s}}$ Fall relation. \\
 
\begin{figure}
\centering
\includegraphics[width=.5\textwidth]{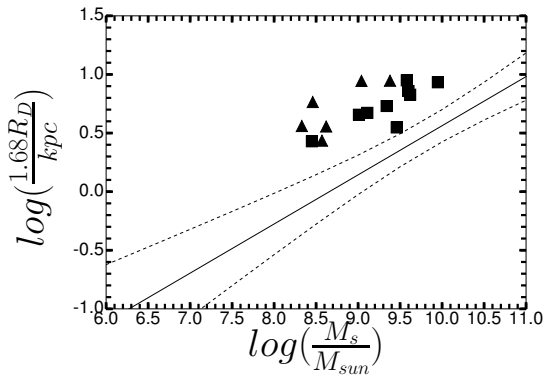}
\caption{Plot of the characteristic radius of an exponential stellar disc $\sim$1.68 $R_D$, $R_D$ being the exponential stellar disc scale length versus the stellar mass M$_{\rm{s}}$. The solid line represents the regression line fit to the ordinary bulgeless spirals and the dotted line the 95.4$\%$ confidence interval of the same.The slope and intercept of the line is given by $0.42\pm{0.11}$ and $-3.63\pm{1.05}$ respectively.  Superposed on the plot are the $j_{\rm{s}}$ versus $V_{\rm{rot}}$ data for the superthins (solid triangles), LSBs (solid squares). The stellar photometry corresponds to the 3.6 $\mu$ band for all the galaxy types except for the LSBs for which it is the B-band. Also, for galaxies 
with two exponential stellar discs, $R_D$ corresponds to the disc scale length of the larger disc.}
\label{fig:1}
\end{figure}

\section{Discussion}

\subsubsection*{Possible origin of discs with high specific angular momenta:} 

\begin{itemize}

\item \textbf{High spin parameter:} During the early phases of their formation and evolution, galaxies acquire their angular momentum by the action of the tidal torque generated by the gravitational field of their global environment (White 1974). The angular momentum thus acquired can be characterized by a dimensionless parameter referred to as the spin parameter $\lambda$, which may be given by $\lambda$ = $\frac{j}{ \sqrt{2} V_{\rm{vir}} R_{\rm{vir}} }$ where $j$ is the specific angular momentum of the disc or the dark matter halo, and R$_{\rm{vir}}$ and V$_{\rm{vir}}$ are the virial radius of the dark matter halo, and its velocity at R$_{\rm{vir}}$ respectively (Bullock et al. 2001). In Table 15, we summarize the 
$\lambda$ parameters for our stellar (${\lambda}_{\rm{s}}$), gas (${\lambda}_{\rm{g}}$) and baryonic discs ${\lambda}_{\rm{b}}$ respectively. We further note here that we have not given the $\lambda$ parameters of several galaxies in this table as we did not have reliable parameters of mass models with an NFW dark matter halo.
Interestingly, we find that both the ${\lambda}_{\rm{s}}$ and ${\lambda}_{\rm{b}}$ values of the superthins are almost an order of magnitude higher than those of the  ordinary discs and most of the LSBs. Investigating the origin of the high values of ${\lambda}_{\rm{s}}$ in superthin galaxy, and its implications for a superthin disc can be a possible future study.\\ 
Besides, earlier studies considered the spin parameters of the disc and the dark matter halo to be equal assuming the conservation of the specific angular momentum of the gas (Fall \& Efstathiou 1980, for example). Therefore, in semi-analytical studies of galaxy formation, the spin parameter of the dark matter halo ${{\lambda}}_{\rm{halo}}$ is often used to predict the size of the stellar disc of the galaxy using the relation R$_e$ = ${\lambda}_{\rm{halo}}$ R$_{\rm{vir}}$ where $R_e$ is the half-mass radius of the galaxy, also assuming the asymptotic rotational velocity V$_{\rm{rot}}$ to be equal to the virial velocity V$_{\rm{vir}}$. 
In Figure 7, we plot of the characteristic radius of an exponential stellar disc 1.68 R$_D$ versus ${\lambda}_{\rm{s}}$ R$_{\rm{vir}}$ for the superthins (solid triangles), LSBs (solid squares) and ordinary discs (solid rhombuses). The large error bars on the LSB data points are due to the uncertainties in the NFW model of their dark matter halos. We find that there is no clear correlation between 1.68 R$_D$ and ${\lambda}_{\rm{s}}$ R$_{\rm{vir}}$, which indicates that the stellar disc size is not strongly governed by their dark matter halo properties. \\ 

\begin{figure}
\centering
\includegraphics[width=0.5\textwidth]{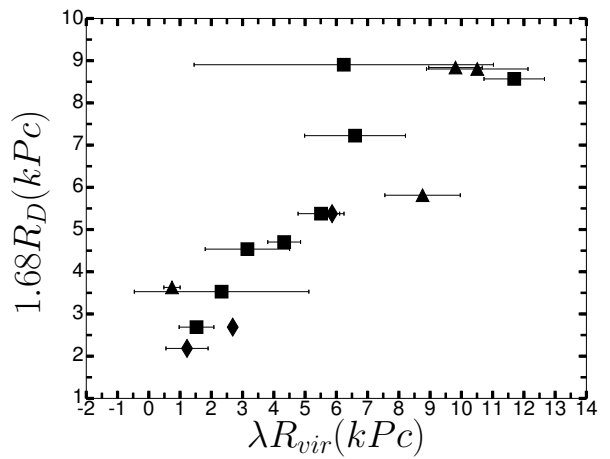}

\caption{Plot of the characteristic radius of an exponential stellar disc 1.68 $R_D$ versus $\lambda$ R$_{\rm{vir}}$ for the superthins (solid triangles), LSBs (solid squares) and ordinary discs (solid rhombuses) where $\lambda$ is the dimensionless spin parameter of the stellar disc given by $\frac{ j_{\rm{stars}} } { {\sqrt(2) V_{\rm{vir}} R_{\rm{vir}} } }$, $V_{\rm{vir}}$ and $R_{\rm{vir}}$ being the virial velocity and virial radius of the galaxy respectively. The stellar photometry corresponds to the 3.6 $\mu$ band for all the galaxy types except for the LSBs for which it is the B-band. Also, for galaxies 
with two exponential stellar discs, $R_D$ corresponds to the disc scale length of the larger disc.}
\label{fig:2}
\end{figure}

\begin{table}
	\centering
	\begin{minipage}{150mm}
	\caption{Dimensionless spin parameter $\lambda$}
	\label{tab:example_table}
	\begin{tabular}{lccc}
		\hline
		Name & ${\lambda}_{s}$ & ${\lambda}_{g}$ & ${\lambda}_{b}$\\
		\hline
		Superthins& & & \\
		\hline
		UGC 7321 & 0.15 $\pm$ 0.01 & 0.27 $\pm$ 0.03 & 0.22 $\pm$ 0.02 \\
		IC 5249  & 0.13 $\pm$ 0.02 & 0.0005 $\pm$ 0.0002 & 0.13 $\pm$ 0.02 \\
		IC 2233  & 0.002$\pm$ 0.001 & 0.003 $\pm$ 0.001 & 0.003 $\pm$ 0.001 \\
		FGC 1540 & 0.18 $\pm$ 0.02 & 0.25 $\pm$ 0.12 & 0.23 $\pm$ 0.09 \\
		\hline
		LSB& & & \\
		\hline
		F563-1  & 0.04 $\pm$ 0.00 & 0.08 $\pm$ 0.07 & 0.07 $\pm$ 0.05 \\
		F568-3  & 0.008 $\pm$ 0.031 & 0.008 $\pm$ 0.032 & 0.008 $\pm$ 0.032 \\
		F579-V1 & 0.13 $\pm$ 0.01 & 0.11 $\pm$ 0.01 & 0.12 $\pm$ 0.01 \\
		F583-1  & 0.01 $\pm$ 0.00 & 0.04 $\pm$ 0.01 & 0.03 $\pm$ 0.01 \\
		F583-4  & 0.03 $\pm$ 0.01 & 0.03 $\pm$ 0.01 & 0.03 $\pm$ 0.01 \\
		F563-V2 & 0.01 $\pm$ 0.01 & 0.02 $\pm$ 0.02 & 0.01 $\pm$ 0.02 \\
		F574-1  & 0.06 $\pm$ 0.01 & 0.05 $\pm$ 0.01 & 0.06 $\pm$ 0.01 \\
		F568-V1 & 0.05 $\pm$ 0.01 & 0.06 $\pm$ 0.01 & 0.05 $\pm$ 0.01 \\
		F568-1  & 0.03 $\pm$ 0.02 & 0.02 $\pm$ 0.01 & 0.02 $\pm$ 0.02 \\
		\hline
		Spirals& & & \\
		\hline
		NGC 2403 & 0.02 $\pm$ 0.00 & 0.05 $\pm$ 0.00 & 0.04 $\pm$ 0.00 \\
		NGC 3198 & 0.04 $\pm$ 0.00 & 0.12 $\pm$ 0.00 & 0.07 $\pm$ 0.00 \\
		NGC 7793 & 0.01 $\pm$ 0.00 & 0.01 $\pm$ 0.00 & 0.01 $\pm$ 0.00 \\
		\hline
		
	\end{tabular}
\end{minipage}
\end{table}

\item \textbf{Galaxy size versus dark matter halo concentration parameter:} Zoom-in cosmological simulation studies by Jiang et al. (2018) have lately revealed that galaxy size is in fact strongly regulated by the concentration parameter $c$ of its dark matter halo according to the equation R$_e$ = $A$ R$_{\rm{vir}}$ where $A$ = 0.02 (c/10)$^{-0.7}$. This implies galaxies with smaller disc size reside in smaller halos and vice-vera. We plotted the characteristic radius of an exponential stellar disc $R_e$ as predicted from cosmological hydrodynamical simulations of Jiang et al. (2018) versus $R_D$, the exponential stellar disc scale length as determined from observations. We obtained the regression line fit to the ordinary bulgeless spirals and the dotted line the 95.4$\%$ confidence interval of the same. Superposed on the plot were the $j_{\rm{s}}$ versus $v_{\rm{rot}}$ data for the superthins (solid triangles) and LSBs (solid squares). However, our results showed that the disc scale length R$_D$ is not driven by the the concentration parameter $c$ of its dark matter halo, and therefore not quite in agreement with the simulation results. \\

\item \textbf{Stellar specific angular momentum versus gas mass fraction:} Recent semi-analytical modelling studies of the inter-relation between the size and the angular momentum of galaxies by Zoldan et al. (2017) have indicated the existence of a strong correlation between the stellar specific angular momentum j$_{\rm{g}}$ and gas mass fraction M$_{\rm{g}}$/{M$_{\rm{b}}$} of galaxies. This could be explained on the basis of the fact that gas-poor galaxies i.e., galaxies with low values of M$_{\rm{g}}$/{M$_{\rm{b}}$} are hosted in larger or more massive halos, and therefore undergo a phase of rapid star formation at early phases of its evolution when the angular momentum acquired by the galaxy is quite low. Gas-rich galaxies, on the other hand, are hosted in smaller halos, and hence undergo star formation at a relatively slower pace until the present epoch when the angular momentum content of the halo is higher.
In Figure 8, we plot the regression line fit to the $j_s$ versus M$_{\rm{g}}$/{M$_{\rm{b}}$} for the data for the ordinary spirals with its 95.4$\%$ confidence interval. Superposed on it are the data points  for the superthins (solid triangles) and LSBs (solid squares). 
We may also note here that the gas mass fraction considered in the above study was M$_{\rm{g}}$/{M$_{\rm{b}}$}  > 0.15, which matches the range of our sample galaxies.
We observe that the positive slope of the regression line complies with the trend predicted by the simulations, and the data for the LSBs are roughly in agreement with it. Interestingly, however, the superthins  seem to be indicating a trend opposite to that predicted by the above study, which is puzzling.  \\

\begin{figure}
\centering
\includegraphics[width = 0.5\textwidth]{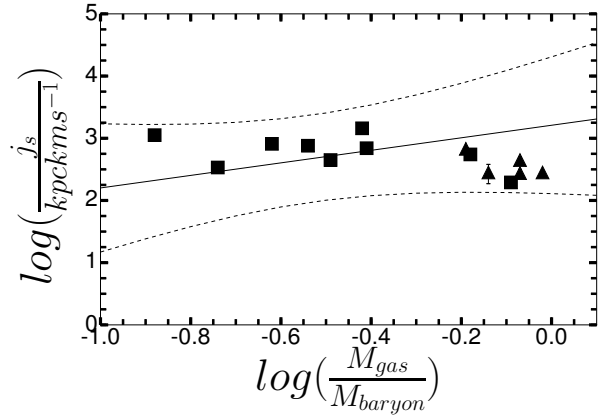}

\caption{Plot of the stellar specific angular momentum $j_{\rm{s}}$ versus the gas mass fraction M$_{\rm{g}}$/{M$_{\rm{b}}$}. The solid line represents the regression line fit to the ordinary bulgeless spirals and the dotted line the 95.4$\%$ confidence interval of the same.The slope and intercept of the line is given by $1.01\pm{1.37}$ and $3.21\pm{0.76}$ respectively.  Superposed on the plot are the data for the superthins (solid triangles) and LSBs (solid squares).}
\label{fig:nfw_5249}
\end{figure} 

\end{itemize}

\subsubsection*{Specific angular momenta of superthins versus those of dwarf-irregulars:}

Recent studies have focused on the comparison of the specific angular momenta of dwarf-irregular galaxies
with those of ordinary disc galaxies (Chowdhury \& Chengalur 2017, Kurapati, Chengalur \& Pustilnik 2018). 
Both dwarf-irregulars and low surface brightness galaxies including superthins constitute the (very) late-type galaxy population.
However, dwarf-irregulars are characterized by a lower dynamical mass as indicated by asymptotic rotational velocity V$_{\rm{rot}}$ values
between 10 - 50 kms$^{-1}$ whereas LSBs have the same between 80 - 120 kms$^{-1}$. In Figure 9 , we present the plot of the specific angular momenta of the stellar discs  j$_{\rm{s}}$ versus the asymptotic rotational velocities V$_{\rm{rot}}$. The solid line represents the regression line fit to the ordinary bulgeless spirals and the dotted line the 95.4$\%$ confidence interval of the same.The slope and intercept of the line is given by $2.51\pm{0.72}$ and $-2.76\pm{1.55}$ respectively. Superposed on the plot are the $j_{\rm{s}}$ versus $V_{\rm{rot}}$ data for the superthins (filled triangles), LSBs (filled squares) and dwarf-irregulars (filled circles). It clearly shows that unlike most of the LSBs and the superthins, the j$_{\rm{s}}$ values of the dwarf-irregulars lie within the 95.4$\%$ confidence band  for that of the ordinary spirals, implying the dwarf-irregulars do not have characteristically higher j$_{\rm{s}}$ values than those of the ordinary spirals. Interestingly, our calculated values of the $\lambda$ parameters of the dwarf-irregulars also came out to be an order of magnitude smaller than those of the superthins and LSBs.

\begin{figure}
\centering
\includegraphics[width = 0.5\textwidth]{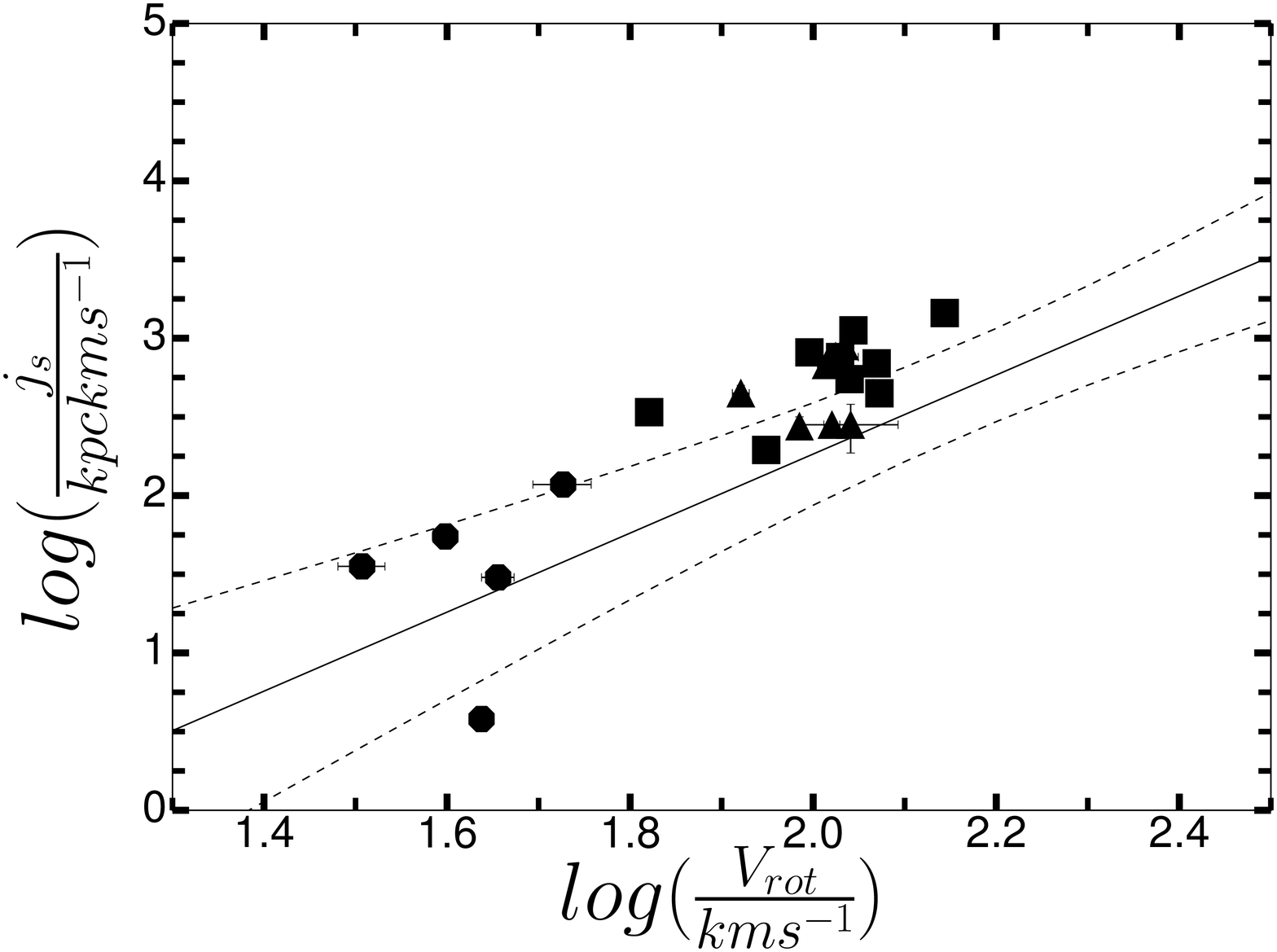} 

\caption{Plot of the specific angular momenta of the stellar discs  j$_{\rm{s}}$ versus the asymptotic rotational velocities V$_{\rm{rot}}$. The solid line represents the regression line fit to the ordinary bulgeless spirals and the dotted line the 95.4$\%$ confidence interval of the same.The slope and intercept of the line is given by $2.51\pm{0.72}$ and $-2.76\pm{1.55}$ respectively}. Superposed on the plot are the $j_{\rm{s}}$ versus $V_{\rm{rot}}$ data for the superthins (filled triangles), LSBs (filled squares) and dwarf-irregulars(filled circles).

\label{fig:nfw_5249}
\end{figure}

\section{Conclusions}

Superthin galaxies are edge-on or nearly edge-on low surface brightness galaxies (LSBs) with strikingly high values of planar-to-vertical axes ratio $b/a$ $\sim$ 10 - 20 with little or no discernable bulge component.
We primarily investigate if the high values of planar-to-vertical axes ratio in superthin galaxies is the reflection of a larger stellar disc size $R_D$  due to a higher value of their stellar specific angular momentum $j_{\rm{s}}$
for a given value of the gravitational potential well, the depth of which may be represented by its aymptotic rotational velocity V$_{\rm{rot}}$ as compared to bulgeless ordinary disc galaxies, as well as to a given stellar mass M$_{\rm{s}}$. Our sample consists of six superthin galaxies for which stellar photometry, atomic hydrogen (HI) surface density and high resolution HI rotation curves were already available in the literature. However the angular momenta of our sample superthin galaxies were not studied earlier. In addition, we also study a sample of nine general LSBs for which all the necessary input parameters were available as well. 

We find that the characteristic sizes of the stellar discs, which is defined as 1.68 R$_D$ where R$_D$ is the exponential stellar disc scale length, for several of the superthins and LSBs and are higher for a given value of $j_{\rm{s}}$ as compared to ordinary bulgeless disc galaxies at high levels of statistical significance. This already hints at the fact that the depth of the gravitational potential wells in which some of the low surface brightness galaxies are hosted are possibly shallower compared to ordinary disc galaxies. In fact, we find that the $j_{\rm{s}}$ values of \emph{only} three out of our six superthins and \emph{only} seven out of our nine LSBs lie above the 95.4 $\%$ confidence band of the $j_{\rm{s}}$ - $V_{\rm{rot}}$ regression line for ordinary bulgeless disc galaxies, $V_{\rm{rot}}$ being the asymptotic rotational velocity while the rest lie within it, although systematically higher than those of the ordinary bulgeless spirals with the same $V_{\rm{rot}}$. \emph{indicating superthin galaxies and LSBs in general may have higher values of $j_{\rm{s}}$  and hence larger disc size R$_D$ compared to ordinary spirals with the same $V_{\rm{rot}}$, which may drive the large planar-to-vertical axes ratios of the stellar discs in superthin galaxies.}

Interestingly, however, we find that the gas specific angular momenta $j_{\rm{g}}$ values of our superthins and LSBs lie within the 95.4 $\%$ confidence band of the $j_{\rm{g}}$ - $V_{\rm{rot}}$ regression line for ordinary bulgeless disc galaxies. \emph{Since the gas discs are the progenitors of the stellar discs, lack of agreement with the $j_{\rm{s}}$ - $V_{\rm{rot}}$   relations inspite of compliance with the $j_{\rm{g}}$ - $V_{\rm{rot}}$   relation is a possible reflection of the characteristically different routes of evolution followed by the ordinary spirals versus the low luminosity galaxies.}

Finally, we also obtain the $j_{\rm{b}}$ - $V_{\rm{rot}}$   for the ordinary spirals and compare them with the data for the superthins and LSBs .
We find that the total baryonic specific angular momenta $j_{\rm{b}}$ values of our superthins and LSBs lie within the 95.4 $\%$ confidence band of the $j_{\rm{b}}$ - $V_{\rm{rot}}$ regression line for ordinary bulgeless disc galaxies, indicating that the gas discs in superthins and LSBs have specific angular momentum $j_{\rm{g}}$ comparable with those of the ordinary bulgeless disc galaxies for a given 
$V_{\rm{rot}}$ or depth of the gravitational potential well. However, in this case, the $j_{\rm{b}}$ values of the superthins and LSBs were found to be systematically higher than those of the ordinary bulgeless disc galaxies. Therefore, the $j_{\rm{b}}$ - $V_{\rm{rot}}$ relation is a reflection of the $j_{\rm{g}}$ - $V_{\rm{rot}}$ in the galaxies. \\

In addition, we investigate the possible origin of a stellar disc with high $j_{\rm{s}}$ values in the superthins and LSBs.
We find that the median spin parameter $\lambda = \frac{ j_{\rm{stars}} } { {\sqrt(2) V_{\rm{vir}} R_{\rm{vir}} } }$, $V_{\rm{vir}}$ and $R_{\rm{vir}}$ being the virial velocity and virial radius of the galaxy respectively,  is 0.13 $\pm$ 0.01 for superthin galaxies which is an order of magnitude higher than those of LSBs and ordinary spirals, which may have important implications for the existence of superthin stellar discs in these low surface
 brightness galaxies.  We also find that the the stellar specific angular momentum $j_{\rm{s}}$  moderately correlates with the gas mass fraction $M_{\rm{g}}$ /$M_{\rm{b}}$ for the ordinary disc galaxies as well as most of the LSBs in agreement with the results of recent numerical studies. Finally, we compare the specific angular momenta of the stellar discs of superthins and LSBs with those of the dwarf-irregulars, another class of gas-rich and dark matter-dominated late-type galaxies but not characterized by superthin discs. We find that for a given value of $V_{\rm{rot}}$, the dwarf irregulars have $j_{\rm{s}}$ values comparable to those of the ordinary discs in contrast to superthins or LSBs.

\section*{Acknowledgements} The authors would like to thank Ms. Sushma Kurapati for providing the data for FGC1540 and also acknowledge DST-INSPIRE Faculty Fellowship (IFA14/PH-101) for supporting this research. We would also like to thank the anonymous referee for the detailed comments which have helped to improve the quality of the paper.

\section*{References}
    
        Banerjee, A. \& Jog, C. J. 2013, MNRAS, 431, 582 \\
	Banerjee A., Matthews L. D., Jog C. J., 2010, NewA, 15, 89 \\
        Barbanis, B., Woltjer, L. 1967, ApJ, 150, 461 \\
        Begum, A. , Chengalur, J.N. 2004, A\&A, 424, 509 \\
	Bizyaev, D. V., Kautsch, S. J., Sotnikova, N. Ya., Reshetnikov, V. P., Mosenkov, A. V., 2017, MNRAS, 465, 3784 \\
	Brandt, J.C. 1960, ApJ, 131, 293 \\
	Bureau, M., Athanassoula, E.  2005, ApJ, 626, 59 \\
	Chowdhury, A., Chengalur, J. N. 2017,  MNRAS, 467, 3856 \\
	Dalcanton, J., Spergel, D. N., Gunn, J. E., Schmidt, M., Schneider, D. P. 1997, AJ, 114, 2178 \\
	Efstathiou G., Lake G., Negroponte J., 1982, MNRAS, 199, 1069 \\
        Garg, P. \& Banerjee, A. 2017, MNRAS, 
        Ghosh S., Jog, C. J. 2014, MNRAS, 439, 929 \\
        Grand, R. J. J., Springel, V., Gomez, F. A., Marinacci, F., Pakmor, R., Campbell, D. J. R., Jenkins, A. 2016, MNRAS, 459, 199 \\
        Karachentsev I., 1989, AJ, 97, 1566 \\
        Karachentseva, V. E., Kudrya, Yu. N., Karachentsev, I. D., Makarov, D. I., Melnyk, O. V.  2016, AstBu, 71, 1 \\
        Kautsch S. J., 2009, PASP, 121, 1297 \\
        Khoperskov A., Bizyaev D., Tiurina N., Butenko M., 2010, Astron. Nachr., 331, 731 \\
        Kurapati, S., Chengalur, J.N., \& Pustilnik, S. 2018, MNRAS, 479, 228 \\
        Matthews, L. D., Uson, Juan M. 2008, ApJ, 688, 237 \\
        Mendelowitz, C. M., Matthews, L. D., Hibbard, J. E., Wilcots, E. M. 2000, BAAS, 32, 1459 \\
        Obreschkow, D., Glazebrook, K.  2014, ApJ, 784, 26 \\
        O'Brien J. C., Freeman, K. C., van der Kruit, P. C. 2010c, A\&A, 515, 62 \\
        Patra, N. N., Banerjee, A., Chengalur, J.N. \& Begum, A. 2014, MNRAS, 479, 5686 \\
        Peebles P. J. E., 1969, ApJ, 155, 393 \\
        Posti, L., Pezzulli, G., Fraternali, F., Di Teodoro, E. M.  2018, MNRAS, 475, 232 \\
        Posti, L., Fraternali, F., di Teodoro, E., Pezzulli, G. 2018, arXiv180404663 \\
        Qu, Y., Di Matteo, P., Lehnert, M. D., van Driel, W. 2011, A\&A, 399, 879 \\
        Rosenbaum, S. D., Krusch, E., Bomans, D. J., Dettmar, R.-J. 2009, A\&A, 504, 807 \\
        Uson, Juan M., Matthews, L. D.	 2003, AJ, 125, 2455 \\
        Velazquez, H., White, S. D. M. 1999, MNRAS, 304, 254 \\
        Walker, I. R., Mihos, J. C., Hernquist, L. 1996, ApJ, 460, 121 \\
        White S. D. M., 1984, ApJ, 286, 38 \\
        Zoldan, A., De Lucia, G., Xie, L., Fontanot, F., Hirschmann, M.  2018, arXiv180308056 \\
        Zschaechner, L. K., Rand, R. J., Heald, G. H., Gentile, G., Kamphuis, P.  2011, ApJ, 740, 35 \\
        
\section*{Appendix}

We study the Fall relation (1983) i.e. the correlation between specific angular momentum and disc mass in the stellar, gas and baryonic discs in our superthin and LSB samples. In Figure 10, we plot the regression line fit to the $j_{\rm{s}}$ - $M_{\rm{s}}$ data along with the 2-$\sigma$ or 95.4 $\%$ confidence band for the ordinary discs,
and superpose the data for the superthins and LSBs and on the same. It clearly shows that for a given stellar mass M$_{\rm{s}}$, superthin (solid triangles) and LSBs (solid triangles) have distinctly higher values of specific angular momenta $j_{\rm{s}}$ compared to ordinary bulgeless spiral galaxies, lying outside the 2-$\sigma$ or 95.4 $\%$ confidence band of the $j_{\rm{s}}$ - $M_{\rm{s}}$ scaling relation for bulgeless disc galaxies. This anomaly with respect to bulgeless disc galaxies may be possibly understood from the origin of the above scaling relation i.e., disc galaxies are rotationally-supported in the plane against 
collapse due to the net gravitational field due to the mass of the stars (M$_{\rm{s}}$), gas (M$_{\rm{g}}$) and the dark matter halo (M$_{\rm{DM}}$) (Obreschkow \& Glazebrook 2014). Therefore $V_{\rm{rot}}$ is equivalent to the total dynamical mass of the galaxy i.e., M$_{\rm{dyn}}$ = M$_{\rm{s}}$ + M$_{\rm{g}}$+ M$_{\rm{DM}}$, and mass of the stars M$_{\rm{s}}$ represents a fraction of M$_{\rm{dyn}}$. 
Therefore, the deviation of the data points corresponding to the superthins and LSBs from the $j_{\rm{s}}$ - $M_{\rm{s}}$ scaling relation obeyed by the ordinary spirals reflects that the stellar mass fraction in the total dynamical mass is significantly smaller than that in ordinary spirals. This is evident given the low star formation rates in the galaxies as is indicated by their low luminosity or low surface brightness nature, as the case may be. 
In fact, we find that late-type galaxies comprising a sample of superthins and  LSBs obey a different scaling 
relation, given by a regression line of slope of 0.41 $\pm$ 0.082, and an intercept of -1.02 $\pm$ 0.749; this is distinct from the regression
 line corresponding to the ordinary spirals, characterized by a slope of 0.87 $\pm$ 0.076 and intercept of -5.85 $\pm$ 0.753. Interestingly, the slope of the regression line obtained for the late-types alone closely matches those for the fundamental angular momentum-mass relation encompassing all galaxy morphologies from dwarf-irregulars to massive spirals (Posti et al. 2018).
 
\begin{figure}
\centering
\includegraphics[width=.5\textwidth]{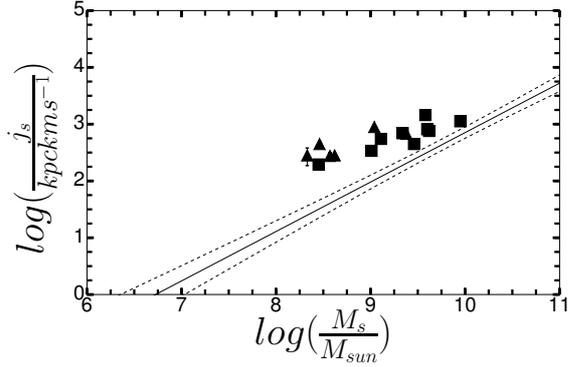}
\caption{Plot of the specific angular momenta of the stellar discs  j$_{\rm{s}}$ versus the stellar mass M$_{\rm{s}}$. The solid line represents the regression line fit to the ordinary bulgeless spirals and the dotted line the 95.4$\%$ confidence interval of the same.The slope and intercept of the line is given by $0.87\pm{0.76}$ and $-5.86\pm{0.75}$ respectively. } Superposed on the plot are the $j_{\rm{s}}$ versus M$_{\rm{s}}$ data for the superthins (solid triangles), LSBs (solid squares)
\label{fig:nfw_5249}
\end{figure}

In Figure 11, we plot the specific angular momenta of gas discs versus gas masses of superthins (solid triangles), LSBs (solid squares) as 
obtained in this paper and compare them with the j$_{\rm{g}}$ - M$_{\rm{g}}$ scaling relation of bulgeless ordinary spirals with its 95.4$\%$ confidence band.
Interestingly, unlike the stellar case, the j$_{\rm{g}}$ values of our sample superthins fall within the 2-$\sigma$ band of the j$_{\rm{g}}$ - M$_{\rm{g}}$ 
scaling relation as the ordinary spirals, and therefore could be said to obey the same scaling relation as the ordinary spirals. 
Arguing along the same lines as in the stellar case, we may say that the gas mass fraction in the total dynamical mass of the galaxy is 
the same in the superthins, LSBs as in the ordinary bulgeless spirals. This is further reflected by the fact that regression lines 
fitted to the respective data sets of the the superthins, the LSBs taken together, and the ordinary bulgeless spirals overlap with each other within error bars; the former has a best-fitting slope of 0.74 $\pm$ 0.077 and an intercept of -3.99 $\pm$  0.705, whereas, for the latter, the values are 0.83 $\pm$ 0.090 and -4.86 $\pm$ 0.89 respectively. This further confirms that the gas mass fraction in the total dynamical mass of the galaxy is roughly constant irrespective of the morphological type of the late-type galaxies.

\begin{figure}

\centering
\includegraphics[width=.5\textwidth]{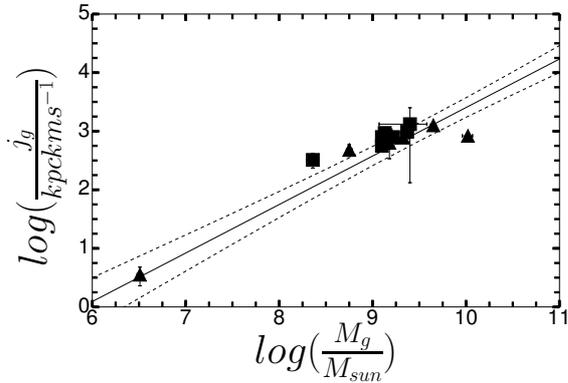}
\caption{Plot of the specific angular momenta of the gas discs  j$_{\rm{g}}$ versus the stellar mass M$_{\rm{g}}$. The solid line represents the regression line fit to the ordinary bulgeless spirals and the dotted line the 95.4$\%$ confidence interval of the same.The slope and intercept of the line is given by $0.83\pm{0.09}$ and $-4.86\pm{0.89}$ respectively. } Superposed on the plot are the $j_{\rm{g}}$ versus M$_{\rm{g}}$ data for the superthins (solid triangles), LSBs (solid squares). 
\label{fig:figure5}
\end{figure}

In Figure 12, we present a regression line fit to the $j_{\rm{b}}$ versus M$_{\rm{b}}$ data for the ordinary bulgeless spirals using the data of Obreschkow \& Glazebrook (2014) along with its 95.4$\%$ confidence band. We also superpose the $j_{\rm{b}}$ versus M$_{\rm{b}}$ data for the superthins (solid triangles) and LSBs (solid squares) on this plot. As in the stellar case disc case, for a given value of M$_{\rm{b}}$,
 all of superthins, LSBs have j$_{\rm{b}}$ values much greater than those of the ordinary discs with the same M$_{\rm{b}}$ value.
 
\begin{figure}

\centering
\includegraphics[width=.5\textwidth]{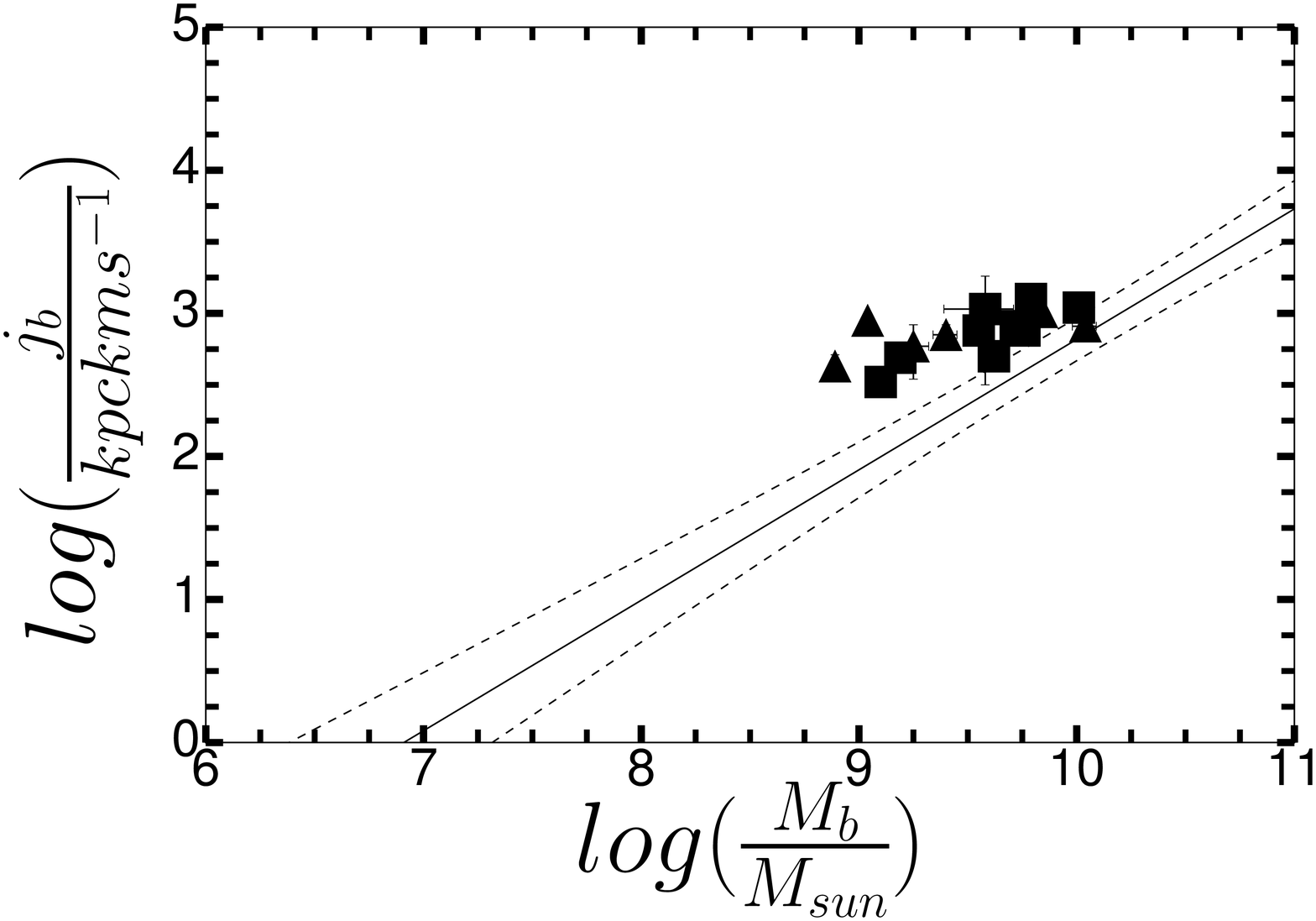}
\caption{Plot of the specific angular momenta of the baryonic discs  j$_{\rm{b}}$ versus the baryonic mass M$_{\rm{b}}$. The solid line represents the regression line fit to the ordinary bulgeless spirals and the dotted line the 95.4$\%$ confidence interval of the same.The slope and intercept of the line is given by $0.91\pm{0.11}$ and $-6.30\pm{1.09}$ respectively. } Superposed on the plot are the $j_{\rm{b}}$ versus M$_{\rm{b}}$ data for the superthins (solid triangles), LSBs (solid squares).
\label{fig:figure4}
\end{figure}

\end{document}